\documentclass[11pt]{article}
\usepackage{epsfig,dsfont}
\usepackage{geometry}
\usepackage{amsmath}
\usepackage{amssymb}
\usepackage[latin1]{inputenc}

\usepackage[normalem]{ulem}
\usepackage{color}

\usepackage{hyperref}

\newcommand{\F}{\mathcal F}
\renewcommand{\S}{\mathcal S}
\newcommand{\D}{\mathcal D}
\newcommand{\C}{\mathcal C}
\newcommand{\be}{\begin{equation}}
\newcommand{\ee}{\end{equation}}
\newcommand{\avg}[1]{\left\langle #1\right\rangle}
\newcommand{\lam}{\lambda}
\newcommand{\U}{\text{U}}
\newcommand{\PSU}{\text{PSU}}
\newcommand{\SU}{\text{SU}}

\def\le{\left}
\def\ri{\right}
\newcommand{\p}{\partial}

\usepackage{cite}

\begin{document}
 
\vspace*{1mm}
\begin{center}
{\LARGE
Phase structure of self-dual lattice gauge theories in 4d
} \\
\vskip18mm
Mariia Anosova$\,^a$, Christof Gattringer$\,^{b,c}$, Nabil Iqbal$\,^d$ and Tin Sulejmanpasic$\,^d$ 
\vskip12mm
$\;^a$ Universit\"at Graz, Institut f\"ur Physik\footnote{Member of NAWI Graz.}, Universit\"atsplatz 5, 8010 Graz, Austria
\vskip1mm
$\;^b$ FWF Austrian Science Fund, Sensengasse 1, 1090 Vienna, Austria
\vskip1mm
$\;^c$ On leave from: Universit\"at Graz, Institut f\"ur Physik, Universit\"atsplatz 5, 8010 Graz, Austria
\vskip1mm
$\;^d$  Department of Mathematical Sciences, Durham University, DH1 3LE Durham, United Kingdom
\end{center}
\vskip15mm

\begin{abstract}
We discuss $\U(1)$ lattice gauge theory models based on a modified Villain formulation of the gauge action, 
which allows coupling to bosonic electric and magnetic matter. The formulation enjoys a duality which maps electric and magnetic sectors into each other.
We propose several generalizations of the model and discuss their 't~Hooft anomalies. A particularly interesting class of theories is the one where electric and 
magnetic matter fields are coupled with identical actions, such that for a particular value of the gauge coupling the theory 
has a self-dual symmetry. The self-dual 
symmetry turns out to be a generator of a group which is a central extension of $\mathbb Z_4$ by the lattice translation symmetry group. The simplest case 
amenable to numerical simulations is the case when there is exactly one electrically and one magnetically charged boson. We discuss the phase structure of 
this theory and the nature of the self-dual symmetry in detail. Using a suitable worldline representation of the system we present the results 
of numerical simulations that support the conjectured phase diagram.
\end{abstract}

\newpage

\tableofcontents

\section{Introduction}

$\U(1)$ gauge theories are widely thought to be effective theories, not only for electrodynamics, but also for many condensed matter systems. In 
four-dimensional spacetime it is the common lore that the U(1) theory of electrodynamics is an effective theory, possibly arising from a non-abelian gauge 
theory in the UV. Such UV completions always require magnetic monopoles, whose masses and effects are fixed by the UV theory. 
Furthermore, it is well known that monopoles 
also emerge in Wilson's lattice regularization of $\U(1)$ gauge theories. In both of these UV completions, the simplest $\U(1)$ gauge theory necessarily has  
minimally charged magnetic monopoles, even when no electric field is present. It is tempting to conclude that any UV completion of a 
$\U(1)$ gauge theory must necessarily contain unit charge magnetic monopoles.

The first hint that this cannot be true in general 
comes from lower dimensional anti-ferromagnetic spin-systems. Such systems are known to be well described by an 
emergent $\U(1)$ gauge theory. The most familiar example is that of an anti-ferromagnetic spin chain that is effectively described by a 1+1d O(3) model 
\cite{Haldane:1982rj}, which in turn flows to an effective 1+1d abelian gauge theory. In two spatial dimensions, anti-ferromagnets have a similar emergent 
$\U(1)$ gauge theory, but now in 2+1 space-time dimensions. Such an effective theory naively has a $\U(1)$ conserved current\footnote{The current is 
$\U(1)$ because the charges are quantized in magnetic flux units.} $j^\mu_T= \frac{1}{2\pi} \epsilon^{\mu\nu\rho}F_{\nu\rho}$\footnote{The current is 
conserved off-shell, so it is commonly called a topological symmetry. Note, however, that this designation depends on the description of the theory. Indeed a 
free $\U(1)$ gauge theory is equivalent to a compact boson, where the topological symmetry becomes a normal Noether-like shift-symmetry of the boson.}, 
where $F_{\mu\nu}$ is the $\U(1)$ field strength tensor. But no such $\U(1)$ symmetry exists in the UV spin system, and so the effective theory must have 
operators that break the $\U(1)$ symmetry explicitly. Such operators are monopoles\footnote{Note that monopoles are space-time localized in 2+1d, i.e., they 
are instantons.}, and they are almost always relevant \cite{Polyakov:1976fu}\footnote{The monopole operators are known/conjectured to be irrelevant at 
certain 3d fixed points (see, e.g., \cite{Senthil:2003eed,Vishwanath:2003yjl,Pufu:2013vpa,Dyer:2015zha,shao2016quantum}).}. This again agrees with the 
lore that UV completions of the $\U(1)$ gauge theory always requires monopoles.

However, there is something different in this setup. Haldane has shown that unit-charge monopoles couple to geometric phases of the underlying spin 
systems \cite{Haldane:1982rj}, and, depending on the underlying lattice symmetries, may obliterate the unit charge monopoles, preserving a discrete 
$\mathbb Z_N$ subgroup of the $\U(1)$-topological magnetic symmetry. Therefore such systems, while not preserving the full $\U(1)$-topological symmetry, 
do preserve a subgroup of it. So at least in such systems monopoles of unit charge, albeit 2+1d monopole-instantons, are not part of an effective theory.

The discussion also makes clear that the absence or presence of dynamical monopoles is a question of symmetry. A theory without dynamical 
monopoles has more symmetries. Indeed in continuous 4d spacetime, the 2-index current 
$J^m_{\mu\nu}=\frac{1}{4\pi}\epsilon_{\mu\nu\rho\sigma}\partial_\mu F^{\rho\sigma}$ is identically conserved by the Bianchi identity in the absence of 
monopoles\footnote{The free 4d $\U(1)$ gauge theory has two 1-form symmetries in the continuum. These follow from the two sets of Maxwell equations 
$\partial_\mu F^{\mu\nu}=0$ and $\partial_\mu \epsilon^{\mu\nu\rho\sigma}F_{\rho\sigma}=0$.}. So an abelian gauge theory a priori allows for more 
symmetries than its non-abelian counterparts.

Furthermore,  free abelian gauge theories have an electric-magnetic self-duality\footnote{Here we refer to self-duality in a broader sense, i.e., that the model 
is mapped to itself, but perhaps with some coupling changed. However, in this work we mostly will 
study the self-duality in a more restrictive sense, i.e., that the 
theory maps exactly to itself under the duality transformation. Self-duality then should be viewed as a symmetry of the theory.}, but such a duality is always 
explicitly broken in continuum descriptions of interacting $\U(1)$ gauge theory because of the inability to couple electric and magnetic matter simultaneously. 
In non-abelian gauge theories the situation is different. In the Coulomb regime of such theories, monopoles can appear as solitons, and ever since Montonen 
and Olive \cite{Montonen:1977sn}, the pursuit of theories with electric and magnetic duality has been of interest, with all known cases being supersymmetric 
theories\footnote{The original conjecture of Montonen and Olive was for a non-supersymmetric Georgi-Glashow model, which, as is now known, does not  
enjoy a self-dual symmetry.}. The non-abelian structure of the theory therefore has two effects. 1: it furnishes a UV completion via asymptotic freedom, and 
2: it allows for a local Lagrangian description of a theory with both electric and magnetic matter.

However, a non-abelian UV completion of an abelian theory, inevitably determines the matter content of the full theory to a large extent. 
For example a common feature of the abelianized non-abelian theory is the presence of magnetic monopoles with unit magnetic 
charge, and the charged W-bosons with unit fixed electric charge.

Still there should be no basic obstacle to formulating $\U(1)$ gauge theories with magnetic and electric matter on equal footing. Indeed such objects are 
mutually local, but the problem is that the continuum Lagrangian cannot be formulated with a local magnetic and electric gauge potential. 
One may therefore refer to such theories (if they exist as continuum theories) as \emph{non-Lagrangian theories}.

However, in \cite{Sulejmanpasic:2019ytl} two of us showed that such theories are possible on the lattice, where a Lagrangian can be formulated in 
such a way that it treats electric and magnetic matter on equal footing\footnote{Similar reasoning was also useful in 2d $\U(1)$ gauge theories with the $\theta$-term \cite{Gattringer:2018dlw,Goschl:2018uma,Sulejmanpasic:2020lyq,Sulejmanpasic:2020ubo}.}. In this formulation monopoles are not artifacts of a lattice theory, but can have an 
action associated with them. Such theories are not only interesting in their own right, but they also may lead to new quantum field theories which are 
beyond the one captured by non-abelian gauge theories. In this paper we continue the line of work initiated in \cite{Sulejmanpasic:2019ytl} (compare also 
\cite{Anosova:2021akr, Anosova:2022cjm}) and study the self-dual U(1) lattice system with a single electrically and a single magnetically charged boson. We 
discuss in detail the possible phase structure of the system and, using a suitable worldline representation, present results of a Monte Carlo simulation  
that allow for checking the conjectured phase diagram. 

The results of the paper are presented in two main sections. Section \ref{sec:generalities} discusses generalities of self-dual modified Villain models 
and is organized as follows: In Subsection~\ref{eq:gauge_partition_sum} we review the pure-gauge construction of \cite{Sulejmanpasic:2019ytl}, and in 
Subsection~\ref{sec:duality_transformation} we discuss the electric-magnetic duality. In Subsection~\ref{sec:selfduality} we present the self-dual symmetry 
and discuss its mixing with lattice translation symmetries. In Subsection~\ref{sec:coupling_to_matter} we introduce matter fields, focusing on bosonic matter 
primarily. In Subsection~\ref{sec:limits} we discuss some computable limits of the scalar self-dual model, while in Subsection~\ref{sec:scalar_field_theory} we 
present the perturbative analysis of scalar models far away from the self-dual point. In Subsection~\ref{sec:generalizations} we discuss larger gauge charge 
and flavor generalizations, and 't Hooft anomalies of the flavor symmetries and 1-form symmetries. We also discuss a non-abelian model with exact electric-
magnetic self-dual symmetry. In Subsection~\ref{sec:phase_diagram} we condense 
all of the previous discussion into a self-dual scalar QED phase diagram.  

Section~\ref{sec:numerics} is devoted to the numerical simulation of one-flavor self-dual scalar QED. 
As a first step we introduce the necessary worldline representation in Subsection~\ref{subsec:worldlines} and then, in 
Subsection~\ref{sec:selfdual}, collect self-dual relations and introduce suitable order parameters which we need for the numerical analysis.
In Subsection~\ref{sec:setup} we discuss the numerical setup and general results of the condensation transition. We show that the theory has 
two transitions, one 1st order and one 2nd order. We analyze these transitions in Subsection~\ref{sec:endpoints}. 

Finally in Section~\ref{sec:conclusion} we conclude and discuss some future prospects, specifically addressing 
the search for the novel interacting conformal fixed points.

\section{Self-dual modified Villain models}\label{sec:generalities}

\subsection{The gauge field partition sum}\label{eq:gauge_partition_sum}

In this subsection we summarize the construction of modified U(1) Villain lattice models which was discussed in detail in 
\cite{Sulejmanpasic:2019ytl}\footnote{This construction was also applied for describing fracton models in 
\cite{Gorantla:2021svj} and for formulating models with non-invertible symmetries \cite{Choi:2021kmx}.}. We consider a 4-d hypercubic lattice $\Lambda$ with 
lattice extents $N_\mu, \mu = 1,2,3,4$ and a  total number of sites $V \equiv N_1 N_2 N_3 N_4$. We may think of $\Lambda$ as the union of 
$\Lambda^{(0)},\Lambda^{(1)},\Lambda^{(2)},\Lambda^{(3)}$ and $\Lambda^{(4)}$, which are the sets of all sites, links, faces, cubes and 
hypercubes\footnote{In general $\Lambda^{(r)}$ is the set of all $r$-cells, where a $0$-cell is a vertex or site, a $1$-cell is a link, edge or bond, a $2$-cell is a 
face, etc.}. The lattice discretization in \cite{Sulejmanpasic:2019ytl} relies on the Villain-like gauge action given by\footnote{Whenever we write a sum 
$\sum_c$ or a product $\prod_c$ over the cells $c$ of the lattice, we always take into account only one, standard orientation of the cell. Moreover, since an 
$r$-cell $c$ of the hypercubic lattice is an $r$-dimensional hypercube uniquely defined by a vertex $x$ and $r$ orthonormal vectors 
$\hat\mu_1,\hat\mu_2,\dots,\hat\mu_r$, we can always write instead of $c$, an ordered multiplet 
$(x,\mu_1\mu_2,\dots\mu_r)$ which uniquely determines $c$, such that the sum $\sum_c$ is defined as $\sum_c\equiv \sum_{x}\sum_{\mu_1<\mu_2<\dots<\mu_r}$, and similarly for the product $\prod_c$.}
\be
S_g [A^e,n] \, = \sum_p \frac{1}{2}\big((dA^e)_p+2\pi n_p\big)^2\;, 
\label{eq:S_def}
\ee
where the sum runs over the plaquettes $p$. $(dA^e)_p$ is the discretized gauge field flux (defined below) around the links $l$ of the plaquette $p$, 
built out of the gauge fields $A^e_l$ living on the links $l$, where the superscript ''$e$'' stands for \emph{electric}, 
emphasizing that this gauge fields naturally couples to \emph{electric matter}.  The $n_p$ are integers living on plaquettes -- the so-called 
\emph{Villain variables}. 

Explicitly, by labeling the square plaquettes on a hypercubic lattice\footnote{While these models can be formulated on arbitrary lattices, they are most natural 
on the hypercubic lattice. The reason is that the hypercubic lattice and its dual lattice are isomorphic, and since, as we will review, the self-dual symmetry 
maps the two into each other, such that the self-dual symmetry is only a genuine symmetry on the hypercubic lattice.} with the lattice site 
$x$ of its root point and two indices $\mu,\nu=1,2,3,4$, we define the exterior lattice derivative operator $d$ by
\be
(dA^e)_{x,\mu\nu} \; = \; A^e_{x+\hat \mu,\nu}-A^e_{x,\nu}-A^e_{x+\hat\nu,\mu}+A^e_{x,\mu}\;,
\ee
where $\hat \mu$ indicates a vector in the direction $\mu$. Using this we define the field strength living on the plaquettes $p$ as
\be
F_{p} \; \equiv \; (dA^e)_p + 2\pi n_p \qquad (\, \Leftrightarrow \, F_{x,\mu \nu} \; \equiv \; (dA^e)_{x,\mu\nu} + 2\pi n_{x,\mu \nu} \, )\;.
\label{eq:F_def}
\ee
We can always restrict $A^e_l$ \footnote{One may be tempted to define a field strength to be just $(dA)_p$. 
However, this is a total derivative, and all fluxes of it over a 
closed 2-cycle on the lattice will vanish. Moreover, we can redundantly allow $A_l$ to take values on the whole real line $\mathbb R$. Then the system enjoys 
a discrete 1-form gauge symmetry, and only the combination $dA_p+2\pi n_p$ is gauge invariant \cite{Sulejmanpasic:2019ytl}.} such 
that $A^e_{l}\in [-\pi,\pi)$. 
The partition function is defined as follows,
\be
Z \; = \; \sum_{\{n\}}\int D[A^e]  \, e^{ \, -\beta \, S_g[A^e,n]}\;,
\label{eq:Z_def}
\ee
where we introduced $\beta=\frac{1}{e^2}$, which plays the role of the inverse electric coupling squared. We also defined some short-hand notation,
\begin{subequations}\label{dAsumn}
\begin{align}
&\int \!\! D[A^e] \; \equiv \; \prod_{l} \int_{-\pi}^\pi \frac{dA_l^e}{2\pi} \; = \; \prod_{x} \prod_{\mu} \int_{-\pi}^\pi \! \! \frac{d A^e_{x,\mu}}{2 \pi} \; , \; \; \\
&\sum_{\{ n \}}  \; \equiv \; \prod_p \sum_{n_p\in\mathbb Z} \; = \; \prod_{x} \prod_{\mu < \nu} \sum_{n_{x,\mu \nu} \in \mathds{Z}} \; .
\end{align}
\end{subequations}
The lattice discretization summarized in \eqref{eq:S_def} -- \eqref{dAsumn} is the well known Villain formulation \cite{Villain:1974ir}
of U(1) lattice gauge theory. 

However, while the theory \eqref{eq:Z_def} has an electric 1-form center symmetry\footnote{The 1-form symmetry can be seen by noting that shifts 
$A_l^e\rightarrow A_l^e+V_l$, with $V_l$ such that $(dV)_p=0$, are a symmetry. Naively, the symmetry group then is $H^1(M,\mathbb R)$, where 
$M$ is the underlying space-time manifold. However, shifts of $A_l^e$ by $2\pi\mathbb Z$ are a gauge symmetry, and hence should not be considered 
a global symmetry. So the group is $H^1(M,\U(1))$. For a 4-torus this is just $\U(1)\times \U(1)\times \U(1)\times \U(1)$, where each 
$\U(1)$ group corresponds to a different cycle of the 4-torus. The operators charged under the various $\U(1)$ parts are the (electric) 
Willson loops wrapping around these cycles. }, it does not
display the  magnetic 1-form symmetry expected in four space-time dimensions. The reason is that if the sum over the $n_p$ is unconstrained, then the 
model has dynamical monopoles, which are identified with the configurations for which $(dn)_c\ne 0$, where $c$ is a 3-cube, labeled by one vertex $x$ and 
there directions $\mu,\nu,\rho$ (see, e.g., \cite{Elitzur:1979uv,Cardy:1981fd,Cardy:1981qy,Sulejmanpasic:2019ytl}). Here $d$ is again the exterior lattice 
derivative, with the  action on 2-forms explicitly defined as
\be
(dn)_{x,\mu\nu\rho} \; = \; n_{x+\hat \rho,\mu\nu}-n_{x,\mu\nu}+n_{x+\hat\mu ,\nu\rho}-n_{x,\nu\rho}+n_{x+\hat\nu,\rho\mu}-n_{x,\rho\mu}\;,
\ee
or, in short hand
\be
(dn)_c \; = \; \sum_{p\in \partial c} n_p\;,
\ee
where $\partial c$ is the boundary of $c$, i.e., the set of all (outward) oriented faces of the cube $c$.

To properly define a free gauge theory, we need to eliminate these monopoles by imposing the constraint 
\be\label{closedness}
(dn)_c \; = \; 0\;, \; \; \forall c\in \Lambda^{(3)}.
\ee 
One way for implementing the constraint is to introduce Lagrange multipliers $A^m_c$ assigned to the cubes $c$ of the lattice $\Lambda$. 
Here the superscript $m$ stands for \emph{magnetic} since, as we will see, the Lagrange multipliers will turn out to be a natural definition of a 
magnetic gauge field. We thus may write the set of constraints \eqref{closedness} as 
\be
\prod_c\int_{-\pi}^{\pi} \!\! \frac{d A^m_{c}}{2\pi} \; e^{\,i\,(dn)_c\, A^m_c} \; = \; \int \!\! D[A^{m}] \; e^{ \, i \sum_{c\in \Lambda^{(3)}}(dn)_c \, A^{m}_c} \; ,
\label{eq:constraint_2}
\ee
where we defined,
\be
\int \!\! D[A^m] \; \equiv \; \prod_c \int_{-\pi}^\pi \!\! \frac{dA^m_c}{2\pi} \; .
\ee

\subsection{Duality transformation}\label{sec:duality_transformation}

Having defined the Villain form of the gauge field Boltzmann factor augmented with the constraint \eqref{eq:constraint_2} we are now
ready to discuss the duality transformation. 
We apply the partial integration formula (see the appendices of \cite{Sulejmanpasic:2019ytl} and \cite{Anosova:2022cjm}) to the exponent
of the constraint \eqref{eq:constraint_2} and find,
\begin{multline}\label{eq:part_int}
\sum_c (dn)_c \, A^m_c \; = \; \sum_{x \atop \mu<\nu<\rho} \!
\big(n_{x+\hat \rho,\mu\nu}-n_{x,\mu\nu}+n_{x+\hat\mu ,\nu\rho}-n_{x,\nu\rho}+n_{x+\hat\nu,\rho\mu}-n_{x,\rho\mu}\big) \, A^m_{x,\mu\nu\rho}
\;=\\= \;  \sum_{x \atop \mu<\nu} \, n_{x,\mu\nu} \, \big(A^m_{x-\hat\rho,\mu\nu\rho}-A^m_{x,\mu\nu\rho}\big) \; = \; - \sum_{p}n_p \, (\partial A)_{p}\;,
\end{multline}
where we defined the \emph{divergence operator} $\partial$ on the lattice,
\be
(\partial A^m)_{x,\mu\nu} \; = \; \sum_{\rho}(A^m_{x,\mu\nu\rho}-A^m_{x-\hat\rho,\mu\nu\rho}) \;,
\ee
which is the discrete version of the continuum divergence $\partial^\rho A^m_{\mu\nu\rho}$.
The lattice divergence operator can also be written as
\be
(\partial A^m)_{p} \; = \, \sum_{c\in \hat\partial p}A^m_{c}\;,
\ee
where $\hat \partial p$ is the coboundary of $p$, i.e., the set of all cubes $c$ whose boundary contains $p$ 
(see again the appendices of \cite{Sulejmanpasic:2019ytl,Anosova:2022cjm} for details and generalizations).

\begin{figure}[t!] 
   \centering
   \includegraphics[width=\textwidth]{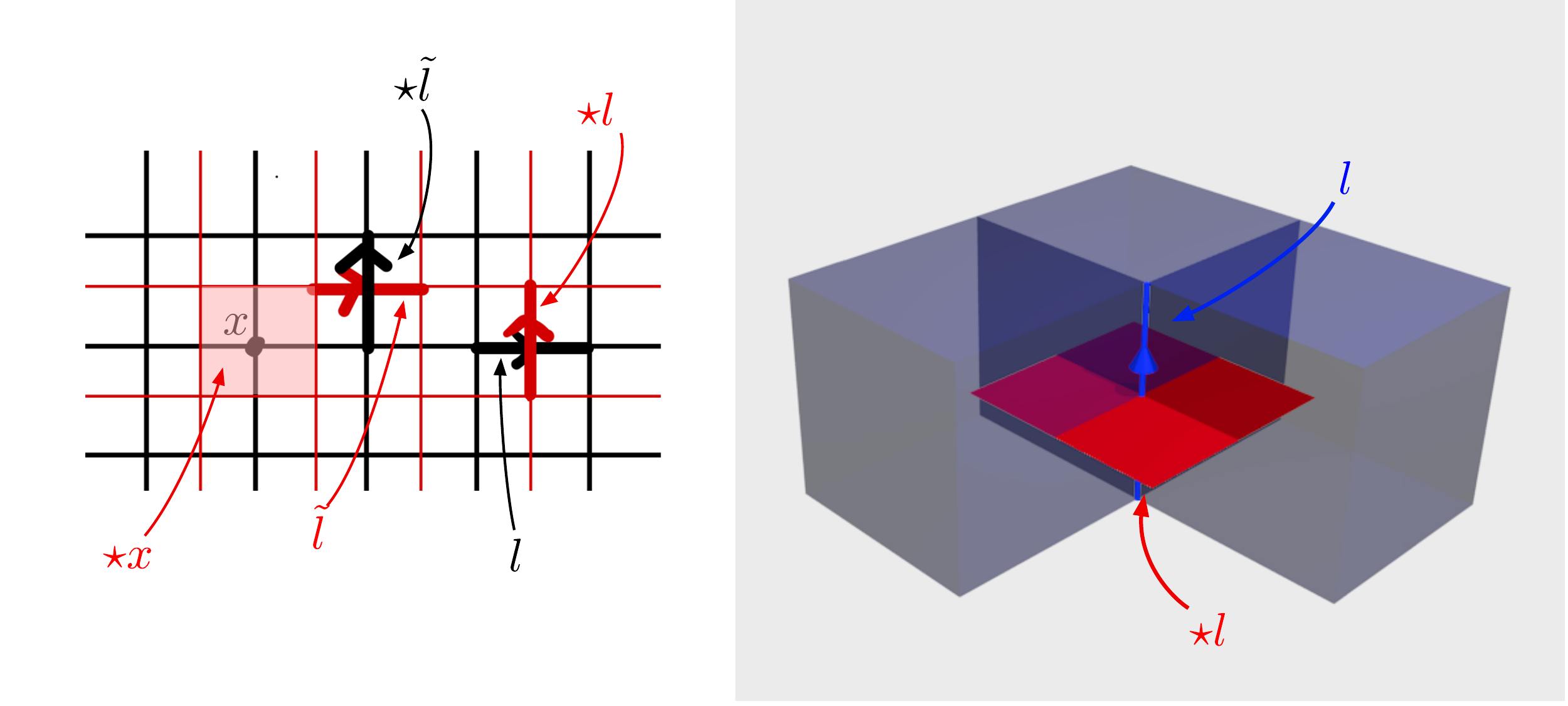} 
   \caption{Examples for the action of the $\star$-operator in 2d (left) and 3d (right).}
   \label{fig:hodge_star}
\end{figure}

Since for the cubic lattice also the dual lattice is cubic, we can define the \emph{Hodge star} (again see, e.g., \cite{Sulejmanpasic:2019ytl}) 
which maps $r$-cells (below denoted as $c^{(r)}_{x,\mu_1\mu_2\cdots \mu_r}$)
of the original lattice $\Lambda$ to $(D-r)$-cells\footnote{A $0$-cell is a vertex, a $1$-cell is a link, a $2$-cell is a face, etc.} of the dual 
lattice $\tilde \Lambda$, where $D$ is the space-time dimension, which here of course is $D = 4$. Analogously one can define a 
$\star$-operator which takes $r$-cells (below denoted as $\tilde c^{(r)}_{\tilde x,\mu_1\mu_2\cdots \mu_r}$) of $\tilde \Lambda$ to $(D-r)$-cells of 
$\Lambda$. We will denote both of these with the same symbol ``$\star$''.  
The star operators are defined as follows
\begin{subequations}\label{app:star}
\begin{align}
&\star c^{(r)}_{x,\mu_1\mu_2\cdots \mu_r}=\sum_{\mu'_{r+1}<\mu'_{r+2}<\dots<\mu'_{D}}\epsilon_{\mu_1\mu_2\dots
\mu_r \mu'_{r+1}\cdots\mu'_D}\tilde c^{(D-r)}_{\tilde x-\hat \mu'_{r+1}-\mu'_{r+2}\cdots-\hat \mu'_{D},\hat\mu'_{r+1}\mu_{r+2}'\cdots 
\mu'_{D}}\;,\label{app:star1}\\
&\star \tilde c^{(r)}_{\tilde x,\mu_1\mu_2\cdots \mu_r}=\sum_{\mu'_{r+1}<\mu'_{r+2}<\dots<\mu'_D}\epsilon_{\mu_1\mu_2\dots
\mu_r \mu'_{r+1}\cdots\mu'_D} c^{(D-r)}_{ x+\hat \mu_{1}+\cdot+\hat\mu_{r},\mu'_{r+1}\mu'_{r+2}\cdots \mu'_{D}}\label{app:star2}\;,
\end{align}
\end{subequations}
where
\be
\tilde x = x+\frac{\hat 1+\hat 2+\dots \hat D}{2}\;,
\ee
is a site on the dual lattice obtained by a translation of $x$. 
Some examples for the action of the $\star$-operator are illustrated in Fig.~\ref{fig:hodge_star}. 

Furthermore, we can define the action of the $\star$-operator on an $r$-form $A_{c^{(r)}}$ living on the $r$-cells $c^{(r)}\in \Lambda^{(r)}$, 
as $(\star A)_{\tilde c^{(D-r)}}\equiv A_{\star \tilde c^{(D-r)}}$. In the case at hand, we have a $3$-form 
$(A^e)_c$ living on cubes $c\in\Lambda^{(3)}$, such that we obtain for $(\star A^m)_{\tilde l}$ living on the dual links $\tilde l$ 
\be
(\tilde A^m)_{\tilde l} \; \equiv \; (\star A^m)_{\tilde l}\equiv (A^{m})_{\star \tilde l}\;.
\ee
We will need the important property that $d\star=\star \partial $, as well as the fact that $\star^2$ acting on a $r$-form $A_{c^{(r)}}$ is the 
identity up to a sign $(-1)^{r(D-r)}$, i.e., $\star^2 A_{c^{(r)}}=(-1)^{r(D-r)}A_{c^{(r)}}$. 

Applying this machinery now to Eq.~\eqref{eq:part_int} we find
\be
\sum_c (dn)_c A^m_c=-\sum_{p}n_p (\partial A^m)_p=-\sum_{p}n_p (\star^2\partial A^m)_{p}=-\sum_{p}n_p(d \tilde A^m)_{\star p}\;,
\ee
such that the partition function is given by
\be\label{eq:part_funct}
Z \; = \; \sum_{\{n\}}\int \! D[A^e] \int \! D[\tilde A^m] \, e^{\, -\frac{\beta}{2}\sum_p \big((dA)_{p}+2\pi n_p\big)^2-i\sum_p n_p (d\tilde A^m)_{\star p}}\;.
\ee
Now we use the Poisson resummation formula
\be\label{eq:poisson}
\sum_n e^{\, -\frac{\beta}{2}(\F+2\pi n)^2-i n \tilde\F} \; = \; 
\frac{1}{\sqrt{2\pi\beta}}\sum_{m}e^{\, -\frac{1}{2(2\pi)^2\beta}(\tilde \F+2\pi m)^2+i \F\tilde \F/(2\pi)+i m \F} ,
\ee
to obtain yet another form of the partition function
\be
Z \; =\; \frac{1}{(2\pi\beta)^{3V}}\sum_{\{m\}}
\int \! D[A^e] \int \! D[A^m] \, e^{\, -\frac{1}{2(2\pi)^2\beta}\sum_p\big((dA^m)_{\star p}+2\pi m_p\big)+i\sum_p m_p (d A^e)_{p}}\;,
\ee
where $V=N_1N_2N_3N_4$ is again the total number of sites of the toroidal 
lattice\footnote{The overall factor $(2\pi\beta)^{-3V}$ is just $1/\sqrt{2\pi\beta}$ from \eqref{eq:poisson} 
to the power of the number of plaquettes of the lattice $6V$.}.
Note that the term $(dA^e)_p (dA^m)_{\star p}$ dropped out because it is 
zero\footnote{This follows from a ``partial integration'' formula, similar to what we used in \eqref{eq:part_int}, and the fact that $d^2=0$.} 
when summed over $p$. By writing $m_p=(\star m)_{\star p}=(\tilde m)_{\star p}$, $(A^e)_p=(A^e)_{\star (\star p)}$ and 
noting that we can replace the sum over $p$ with the sum over $\tilde p=\star p$. Therefore the dual form of the partition function is given by
\be\label{eq:Zdual}
Z \; = \; \frac{1}{(2\pi\beta)^{3V}}\sum_{\{m\}}\int \! D[A^e] \int \! D[\tilde A^m] \, 
e^{-\frac{1}{2(2\pi)^2\beta}\sum_{\tilde p}\big((dA^m)_{\tilde p}+2\pi \tilde m_{\tilde p}\big)^2+i\sum_p \tilde m_{\tilde p} (d A^e)_{\star  \tilde p}}\;.
\ee
This is our final expression for the duality transformation\footnote{This duality was referred to as Kramers-Wannier duality in 
\cite{Sulejmanpasic:2019ytl}, which is a bit imprecise because the original Kramers-Wannier duality does not in fact map a theory 
back to itself, but maps it back to the same theory coupled to a TQFT \cite{Kapustin:2014gua}. In contrast the electric-magnetic 
duality we discuss here is an exact duality. It can be converted to a Kramers-Wannier type of duality by gauging one of the 1-form symmetries. 
This was discussed in \cite{Choi:2021kmx} (see also \cite{Kaidi:2021xfk}).}. Notice that it is almost identical (up to details we will discuss soon) 
to \eqref{eq:part_funct}, with $\beta\rightarrow \tilde\beta =\frac{1}{(2\pi)^2\beta}$. Moreover when $\beta=\beta^*=\frac{1}{2\pi}$ the theory is 
self-dual, and hence has an enhanced symmetry. The self-duality in fact does not square to unity, but to charge-conjugation $C$. 
The reason for this is that in \eqref{eq:Zdual} the phase term has a different sign than in \eqref{eq:part_funct}, which means that 
the self-duality requires not just exchanging the gauge field $A^e$ with $\tilde A^m$, but also flipping the sign of one of them, 
hence squaring to a pure charge conjugation. However, even this is too naive in our lattice theory, as the exchange of the two 
gauge fields also requires shifting the lattice to the dual lattice, and we will see that the self-dual transformation squares to a 
charge conjugation $C$ and an overall lattice shift. We discuss this in detail below.

\subsection{Self-duality and self-dual symmetry}
\label{sec:selfduality}

Comparing the dual form \eqref{eq:Zdual} with the original partition sum \eqref{eq:part_funct} we see that the change 
$(A^e,\tilde A^m)\rightarrow (-\tilde A^m,A^e), 2\pi\beta\rightarrow /(2\pi\beta)$ switches the original partition function to the structure of the 
dual one. However, we cannot simply replace $(A^e,\tilde A^m)$ with $(-\tilde A^m,A^e)$, because $A^e$ lives on the original lattice $\Lambda$, while 
$\tilde A^m$ lives on the dual lattice $\tilde \Lambda$. So we must define the dual transformation such that it also incorporates the map from $\Lambda$ to 
$\tilde \Lambda$. We accomplish this by defining the translation operator $T$, that shifts the lattice by the vector $\hat s= \frac{\hat 1+\hat 2+\dots \hat D}{2}$,
i.e., it shifts from $\Lambda$ to $\tilde \Lambda$. Thus the operator $T$ translates the $r$-cell of the lattice $\Lambda$ to the $r$-cell of the dual 
lattice $\tilde \Lambda$ as follows
\be
Tc^{(r)}_{x,\mu_1\mu_2\dots\mu_r} \; = \; \tilde c^{(r)}_{x+\hat s,\mu_1\mu_2\dots \mu_r}\;.
\ee
Consequently, if we perform the replacement
\begin{equation}\label{eq:selfdual_transformations}
\begin{array}{lcl}
A_l^e&\rightarrow &-\tilde A^m_{T(l)}\;,\\
\tilde A_{\tilde l}^{m} &\rightarrow &A^{e}_{T(\tilde l)}\;,\\
\tilde m_{\tilde p}&\rightarrow &n_{T(\tilde p)}\;,
\end{array}
\end{equation}
in \eqref{eq:Zdual}, the same form as in \eqref{eq:part_funct} is obtained\footnote{Note that $T\star=\star T$ 
as can be checked from the expressions \eqref{app:star}.}, up to an overall constant and of course the change of 
coupling $\beta\rightarrow \tilde \beta = \frac{1}{(2\pi)^2\beta}$. 
Thus we find
\be
Z(\beta) \; = \; (2\pi\beta)^{-3V} Z\left(\frac{1}{\beta(2\pi)^2}\right)\;,
\ee
which we can also write in a more symmetric way as follows,
\be
(2\pi\beta)^{\frac{3V}{2}}Z(\beta) \; = \;  (2\pi\tilde\beta)^{\frac{3V}{2}}Z(\tilde\beta) \; \quad \mbox{with} \quad \tilde\beta= \frac{1}{4\pi^2\beta} \; .
\ee
This relation for the partition sum generates relations for observables: taking the logarithm of both sides 
and differentiating with respect to $\beta$, we obtain a relation between the expectations values of the field strength squared as
\begin{eqnarray}
\langle F^2 \rangle_\beta & \equiv & - \, \frac{1}{3V} \frac{\partial}{\partial \beta} \ln Z(\beta) \; = \;  
- \, \frac{1}{3V} \frac{\partial}{\partial \beta}
\ln \left( (2\pi \beta)^{-3V} Z\big (\widetilde{\beta}) \right) 
\nonumber \\
& = & 
\frac{1}{\beta} \, - \, \frac{1}{3V} \left( \frac{\partial}{\partial \widetilde{ \beta}}  \ln Z(\widetilde{\beta}) \right) \frac{ d \widetilde{\beta}}{d \beta}
\; = \; \frac{1}{\beta} \; - \; \langle F^2 \rangle_{\widetilde{\beta}} \; \frac{1}{4 \pi^2 \beta^2} \; ,
\end{eqnarray}
where we defined $F^2 \equiv \sum_{x,\mu < \nu} (F_{x,\mu \nu})^{2}/(6V) = S_g[A^e\!,\,n] / (3V) $. The subscripts $\beta$  and $\tilde\beta$ attached 
to $\avg{F^2}$ indicate the coupling the respective vevs  are computed with. The relation can be rearranged in a more symmetric 
form as \cite{Anosova:2022cjm}
\be\label{sumruleF2}
\beta\avg{F^2}_\beta+\tilde\beta\avg{F^2}_{\tilde\beta} \, = \, 1\;.
\ee
Another derivative with respect to $\beta$ generates a duality relation for the corresponding susceptibilities,
\begin{equation}
\beta \, \langle F^2 \rangle_\beta \; - \; \beta^2 \, \chi_\beta 
\; = \; \widetilde{\beta} \, \langle F^2 \rangle_{\widetilde{\beta}} \; - \; {\widetilde{\beta}}^2 \,  \chi_{\widetilde{\beta}} \; ,
\label{sumrulechi}
\end{equation}
with
\begin{equation}
\chi_\beta \; \equiv \;  -\frac{\partial}{\partial \beta}   \langle F^2 \rangle_\beta  \; = \;    
3V \left \langle (F^2 - \langle F^2 \rangle_\beta)^2 \right \rangle_\beta \; .
\end{equation}

We conclude this subsection by identifying the \emph{self-dual symmetry}.  
When $\beta=\tilde\beta=\frac{1}{4\pi^2\beta}$, the duality maps our theory to itself, and hence becomes a genuine symmetry. 
By virtue of \eqref{eq:selfdual_transformations}, we define the self-duality operator $\S$ to act on gauge fields as follows (and of course 
$\beta \rightarrow \tilde \beta$)
\begin{equation}
\S=\left\{\begin{array}{lcl}
A_l^e&\rightarrow &-\tilde A^m_{T(l)}\;,\\
\tilde A_{\tilde l}^{m} &\rightarrow &A^{e}_{T(\tilde l)}\;.\\
\end{array}\right.
\end{equation}
The square of the transformation is given by (note that $\tilde{\tilde \beta} = \beta$)
\begin{equation}
\S^2:
\left\{\begin{array}{lcl}
A_l^e&\rightarrow &- A^e_{T^2(l)}\;,\\
\tilde A_{\tilde l}^{m} &\rightarrow &-\tilde A^{m}_{T^2(\tilde l)}\;,
\end{array}\right.
\end{equation}
i.e., it acts as charge conjugation and a diagonal translation of the lattice in all directions by one lattice unit, which we will call $\D$. 
The translation $\D$ forms a group $\mathbb Z$ on an infinite lattice, while on our finite periodic lattice we find
$\D^N=\mathbb I$ for some integer $N$, so that $\D$ generates a $\mathbb Z_N$ subgroup, which we will call 
$Z$. The duality transformation $\S$, the charge conjugation $\C$ and diagonal shifts $\D$ furnish a group $G$, 
which satisfies the following identities
\begin{align}
&\S^2=\C\D\;,\\
&\C^2=\mathbb I \; .
\end{align}
The generator $\D$ commutes with $\S$ and $\C$ and so $Z$ is the center of $G$. 
Furthermore, $G/Z\cong \mathbb Z_4$. Indeed $\S^4=\D^2$ is a pure center element, and so is equivalent 
to the identity element in $G/Z$. One way to phrase this is to say that $G$ is a central extension of $\mathbb Z_4$ by $\mathbb Z$. 

The fact that this algebra involves the lattice shift $\D$ makes clear that the self-dual symmetry does not act in an {\it on-site} manner; 
its action necessarily involves lattice translations. This is consistent with the no-go theorem of \cite{Kravec:2013pua}, 
which argues that any theory with electric-magnetic self-duality cannot be regularized in a manner which realizes the symmetry in an on-site fashion. 
 
If we assume, however, that the lattice symmetries are unbroken by the vacuum of the theory, then the vacuum can at most be acted on by 
$G/ Z \cong \mathbb Z_4$. This $\mathbb Z_4$ group is generated by a self-duality generator $\S$ for which $\S^2=\C$ is the charge-conjugation. 
If we further assume that charge conjugation is unlikely to be broken in the vacuum (below we justify this with numerical results for the system we studied), 
we then conclude that the vacuum will transform at most  under $\mathbb Z_2\cong \mathbb Z_4/\mathbb Z_2$. 
If this happens we will say that self-dual symmetry is spontaneously broken.

\subsection{Coupling electric and magnetic matter}
\label{sec:coupling_to_matter}

We now couple \emph{electrically charged matter} $\phi^e$ to the electric gauge field $A^e$ 
and \emph{magnetically charged matter} $\tilde\phi^m$   to the magnetic gauge field
$\tilde A^m$. We write the partition function with gauge fields coupling to electric and magnetic matter fields in the form
\begin{equation}
Z(\beta, J_e,J_m) \; \equiv \; \int \!\! D[A^e] \int \!\! D[A^m] \; B_\beta[A^e,
\tilde A^m] \;
Z [A^e,J_e] \; \widetilde{Z}\big[\widetilde{A}^m,J_m\big] \; ,
\label{Zfull}
\end{equation}
where $B_\beta[A^e,\tilde A^m]$ is the weight of the free gauge theory derived in \eqref{eq:part_funct}, i.e.,
\be\label{boltzmann_both}
B_\beta[A^e,\tilde A^m]=e^{\, -\frac{\beta}{2}\sum_p\big((dA)_{p}+2\pi n_p\big)^2-i\sum_p n_p (d\tilde A^m)_{\star p}}\;.
\ee
In \eqref{Zfull} we introduced the partition sums $Z$ and $\tilde Z$ for  electric and magnetic matter (defined below),
 which also depend on two coupling parameters 
$J_e$ and $J_m$ for the electric and magnetic matter fields. 
The electrically charged matter fields $\phi^e_x \in\U(1)$ we parameterize as $\phi^e_x = e^{\, i \varphi_x^e}$ with $\varphi_x^e \in [-\pi, \pi)$. 
They couple to the electric background gauge field $A^e$ via the partition sum
\begin{eqnarray}
Z[A^e,J_e] & \equiv &  \int \!\! D[\phi^e] \, e^{ \, J_e S_e[\phi^e,\,A^e]} \; , \; \; 
\int \!\! D[\phi^e] \; \equiv \; \prod_x \int_{-\pi}^\pi \frac{d \varphi_x^e}{2 \pi} \; ,
\label{Zelectric}
\\
S_e[\phi^e,A^e] & \equiv & \frac{1}{2}  \sum_{x,\mu} 
\left[ \phi_x^{e\, *} \,  e^{i A_{x,\mu}^e} \, {\phi_{x+\hat{\mu}}^e} + c.c. \right] 
\; = \; 
\sum_{x,\mu} \cos \big( \varphi_{x+\hat{\mu}}^e - \varphi_x^e + A_{x,\mu}^e \big) \; .
\end{eqnarray}
The magnetically charged scalar field $\widetilde{\phi}_{\tilde{x}}^m \in \U(1)$, which
we parameterize as  $\widetilde{\phi}_{\tilde{x}}^{m} = e^{\, i \widetilde{\varphi}_{\tilde{x}}^{\, m}}$ 
with $\widetilde{\varphi}_{\tilde{x}}^{\, m} \in [-\pi, \pi)$,
lives on the sites $\tilde{x}$ of the dual lattice and couples to 
the magnetic gauge field $\widetilde{A}_{\tilde{x},\mu}^m$ on the links of the dual lattice. 
The corresponding partition sum has the same form 
as the partition sum (\ref{Zelectric}) for the electric matter but for the magnetic matter is defined entirely on the dual lattice:
\begin{eqnarray}
\widetilde{Z}\big[\widetilde{A}^m,J_m\big] & \equiv &  \int \!\! D\big[\widetilde{\phi}^m\big] \, 
e^{\, J_m \widetilde{S}_m \big[\widetilde{\phi}^m,\, \widetilde{A}^m \big]} \; , \; \; 
\int \!\! D \big[\widetilde{\phi}^m\big] \; \equiv \; \prod_{\tilde{x}} \int_{-\pi}^\pi 
\frac{d \widetilde{\varphi}_{\tilde{x}}^{\,m}}{2 \pi} \; ,
\label{Zmagnetic}
\\
\widetilde{S}_m\big[\widetilde{\phi}^m,\widetilde{A}^m\big] & \equiv &  \frac{1}{2}  
\sum_{\tilde{x},\mu} \left[  
\widetilde{\phi}_{\tilde{x}}^{m \, *} \,  e^{i \widetilde{A}_{\tilde{x},\mu}^m}  \, \widetilde{\phi}_{\tilde{x}+\hat{\mu}}^m 
+ c.c. \right] \; = \;   
\sum_{\tilde{x},\mu} \cos \big( \widetilde{\varphi}_{\tilde{x}+\hat{\mu}}^{\,m}
- \widetilde{\varphi}_{\tilde{x}}^{\,m}  + \widetilde{A}_{\tilde{x},\mu}^{\,m} \big) 
\; .
\end{eqnarray}
Here we have coupled electric and magnetic matter using $\U(1)$-valued matter fields, but it is straightforward to generalize this
construction to complex-valued bosonic matter or also to fermionic fields  \cite{Sulejmanpasic:2019ytl}.

Self-duality of the full theory essentially follows from the self-duality of the pure gauge theory we already discussed above, combined with the
interchange of electric and magnetic matter (see \cite{Sulejmanpasic:2019ytl,Anosova:2022cjm} for a more detailed discussion). 
The corresponding self-duality relation for the partition function is given by
\begin{eqnarray}
&& Z(\beta, J_e, J_m) \; = \; \left(\!\frac{1}{2\pi \beta}\!\right)^{\!\!3V}  \,
Z\big(\widetilde{\beta}, \widetilde{J}_e, \widetilde{J}_m\big) \; ,
\label{Zfullduality} \\
&& \mbox{with} \qquad \widetilde{\beta} \; = \; \frac{1}{4\pi^2 \beta} \; , \quad 
\widetilde{J}_e \; = \; J_m  \; , \quad 
\widetilde{J}_m  \; = \; J_e  \; .
\nonumber
\end{eqnarray}
Again we can generate self-duality relations for observables by evaluating derivatives of $\ln Z$ 
with respect to the couplings. 
The pure gauge theory sum rules (\ref{sumruleF2}) and (\ref{sumrulechi}) generalize to,
\begin{equation}
\beta \, \langle F^2 \rangle_{\beta, J_e, J_m} \; + \; \widetilde{\beta} \, 
\langle F^2 \rangle_{\widetilde{\beta}, \widetilde{J}_e, \widetilde{J}_m} \;  = \; 1\; ,
\label{sumruleF2_qed}
\end{equation}
and
\begin{equation}
\beta \, \langle F^2 \rangle_{\beta, J_e, J_m} \; - \; \beta^2 \, \chi_{\beta, J_e, J_m}
\; = \; \widetilde{\beta} \, \langle F^2 \rangle_{\widetilde{\beta}, \widetilde{J}_e, \widetilde{J}_m}  \; - \; 
{\widetilde{\beta}}^2 \,  \chi_{\widetilde{\beta}, \widetilde{J}_e, \widetilde{J}_m} \; .
\label{sumrulechi_qed}
\end{equation}
Derivatives with respect to $J_e$ and $J_m$ generate field expectation values for the electric and magnetic matter fields. 
Exploring the duality relation (\ref{Zfullduality}) one finds the following self-duality relation for the electric and the magnetic action densities $s_e \equiv S_e/4V$ and $\widetilde{s}_m \equiv \widetilde{S}_m/4V$:
\begin{equation}
\left\langle s_e \right\rangle_{\beta, J_e, J_m} \; \equiv \; 
\frac{1}{4V} \frac{\partial }{\partial J_e} \, \ln Z (\beta, J_e, J_m) \; = \; 
 \left\langle \widetilde{s}_m \right\rangle_{\widetilde{\beta}, \widetilde{J}_e,\widetilde{J}_m} \; .
\label{duality_phi2}
\end{equation}
According to the self-duality relation (\ref{duality_phi2}), the electric and magnetic field expectation 
values are converted into each other when 
changing from weak to strong coupling and simultaneously interchanging the electric and 
magnetic coupling parameters. 

\subsection{Computable self-dual limits}\label{sec:limits}

Let us now discuss various limits of the model. First we consider the limit when $J_e=J_m=J\rightarrow0$. 
In this case the model is a pure gauge theory model. Moreover the model has so-called $1$-form magnetic and electric symmetries, 
which can be seen from \eqref{boltzmann_both} by taking $A^e\rightarrow A^e_l+\delta_l$ with $\delta^e_l$ such that $(d\delta^e)_l=0$, 
and similarly for $\tilde A^m_{\tilde l} \rightarrow \tilde A^m_{\tilde l}+\tilde \delta^m_{\tilde l}$ with $(d\delta^e)_{\tilde l}=0$. These 1-form symmetries 
are continuous $\U(1)$ symmetries in the case of a free theory. They act non-trivially on Wilson loops $W^e(C)$ and 't~Hooft loops 
$W^m(\tilde C)$ defined as
\be
W^e(C)= e^{i\sum_{l\in C}A^e_l}\;, \qquad W^m(\tilde C)=e^{i\sum_{l\in \tilde C}\tilde A^m_{\tilde l}} \;,
\ee
where $C$ and $\tilde C$ are closed paths on the lattice and the dual lattice, respectively. 

Moreover the model has a 't~Hooft anomaly between these two symmetries, which in a way guarantees that the photon is exactly massless 
in a free theory, as the anomaly is generically saturated by breaking either the electric or the magnetic 1-form symmetry, leaving a 
goldstone boson -- the photon \cite{Gaiotto:2014kfa}. We discuss this anomaly, or rather the anomaly between the discrete subgroups 
of the electric and magnetic 1-form symmetries, in Sec.~\ref{sec:generalizations}. We can see the presence of this massless phase 
explicitly by integrating out the field $\tilde A^m$. We solve the constraint $(dn)_c=0$ by setting 
\be
n_p= (dk)_p\;,
\ee
where $k_l$ is an integer living on links. This ansatz is correct only locally and not fully general on our compact lattice, such that 
we now formally take the lattice to be infinite to illustrate our point\footnote{Alternatively we can take the $k_l$ to have general boundary 
conditions on our toroidal lattice.}. Now we can remove $k_l$ by a shift of gauge fields\footnote{In order to do this we must a priori take 
$A_l\in \mathbb R$, which we can always do \cite{Sulejmanpasic:2019ytl}.} $A_l^e\rightarrow A_l^e-k_l$ and obtain the action
\be\label{eq:noncompact_action}
\beta S_g=\sum_{p}\frac{\beta}{2}(dA^e)_p^2\;,
\ee
which describes free photons and has no phase transition as a function of coupling because now the coupling can be absorbed 
in the field redefinition\footnote{It is tempting to say that the action \eqref{eq:noncompact_action} describes a non-compact gauge field. 
However, this is not quite right, as the choice of gauge was not possible to perform on a compact manifold, nor is it compatible with the 
insertion of a 't~Hooft line -- two distinguishing features of the compact gauge theory. Rather this should be thought of as a compact 
gauge theory where there are no dynamical monopoles, i.e., they have been suppressed by the constraint on the Villain variables $n_p$. 
Such a theory is locally the same as the non-compact $\mathbb R$-gauge theory, but not globally.} $A_l^e\rightarrow A_l^e/\sqrt{\beta}$. 

Now let us discuss the opposite limit and take $J^m=J^e=J\rightarrow \infty$ in \eqref{Zelectric} and \eqref{Zmagnetic}. 
In this case the matter contribution becomes dominant and will pin the gauge fields to values such that the phase difference 
between sites carries no action cost. This effectively imposes the constraint that
\be
\begin{array}{l}
A^e_l=2\pi s^e_l\\
A^m_l=2\pi s^m_l 
\end{array}\;,\qquad s^e_l,s^m_l\in\mathbb Z\;.
\ee
Notice that the phase factor in the action \eqref{boltzmann_both} drops out and our total action is just
\be
S_{tot}= \frac{\beta}{2}(2\pi)^2\sum_p((ds^e)_p+n_p)^2\;.
\ee
Moreover the Villain variables $n_p$-s are now unconstrained and we can shift them 
$n_p \rightarrow n_p-(ds^e)_p$ to eliminate $s^e_l$ completely. Finally our partition function is given by a product of the plaquette contributions
\be
Z_{J=\infty}\equiv\lim_{\substack{J^e\rightarrow\infty\\J^m\rightarrow\infty}}Z(\beta,J^e,J^m)= \left(\sum_{n} e^{-\frac{\beta}{2\pi^2}n^2}\right)^{6V}\;,
\ee
where $6V$ is the total number of plaquettes. We can express the result via the elliptic function $\theta_3$ and obtain
\be
Z_{J=\infty}=\left(\theta_3(0,e^{-\frac{\beta}{2\pi^2}})\right)^{6V}\;.
\ee
This theory is clearly trivially gapped, and has no phase transition as a function of $\beta$. The fact that the theory is now featureless 
and that there is a distinction between electric and magnetic condensation is the Shenker-Fradkin continuity\cite{Fradkin:1978dv}. 

What about intermediate values $J_e=J_m=J$? This regime is generally strongly coupled, and we will have 
to resort to lattice simulations to answer the question fully, but before we do so, let us perform a qualitative analysis to see what we can expect. 

\subsection{Field-theoretical description away from self-duality}\label{sec:scalar_field_theory}

Let us first consider moving away from self-duality by studying the regime where the inverse gauge coupling $\beta$ is large. 
The electric matter is then weakly coupled, but the magnetic matter is strongly coupled. 
\footnote{We note that by performing the duality transformation \eqref{Zfullduality}, the same analysis describes the large $\tilde{\beta}$ 
regime, and thus may be re-interpreted as the confinement transition associated with the condensation of magnetic charges. 
To avoid confusion we will use terminology appropriate to the Higgs transition.} So we expect the magnetic matter to get a large mass and 
decouple from the system, and the system is described by the condensation of electrically charged matter. Altering $J$ is then related to 
the change of the mass squared parameter $m^2$ of the scalar. The limit where $J\rightarrow \infty$ corresponds to the deep Higgs phase where 
$m^2\rightarrow -\infty$ causes the scalar to condense. On the other hand $J\rightarrow 0$ corresponds to the scalar decoupling limit $m^2\rightarrow \infty$. 

The transition between the two regimes is expected around $m^2\sim 0$. If the mass is low enough, we expect a description in terms of 
continuum scalar QED. This is an extremely well-studied system, where the usual Coleman-Weinberg analysis \cite{Coleman:1973jx} shows 
that the one-loop effective potential for the scalar generically has a minimum away from the origin, and thus the transition is expected to be first order. 

This can also be understood from a renormalization group analysis, as we briefly review below.  We present the argument for 
$N_f^e$ scalar flavors, where the $e$ in the superscript stands for ``electric'' as the matter is electrically coupled. In this section 
we will omit the superscript $e$ (i.e., $N_f = N_f^{e}$) as we are considering only electrically coupled matter, but in later sections 
we will sometimes consider $N_f^{e}$ electrically coupled flavors and $N_f^{m}$ magnetically coupled flavors. 

We now want to discuss the order of the transition, which changes for $N_f$ sufficiently large. The continuum action describing the interaction of electrically charged matter with the photon takes the form
\be
S[\phi,A] = \int d^4x \le(\frac{1}{4e^2} F^2 + \sum_{i=1}^{N_f} \le(|\p \phi_i|^2 + m^2 |\phi_i|^2\ri) + 
\frac{\lambda}{4} \le(\sum_{i=1}^{N_f}|\phi_i|^2\ri)^{\!2} \,\ri) \; .
\label{actionNfscalars} 
\ee 
There are three couplings which are of interest: the electric coupling squared $e^2$, the mass squared $m^2$ and the $\phi^4$ coupling $\lambda$. When 
$m^2>0$, the relevance of this coupling drives the system to a free photon phase. When $m^2<0$ the system is likewise driven to a scalar 
condensed phase. We thus have to understand what happens exactly at $m^2=0$. At this point $e^2$ and $\lambda$ are marginally irrelevant
and an analysis of the beta functions shows that they flow logarithmically towards smaller values of the couplings $e^2, \lambda$. 

The precise character of the transition now turns out to depend on $N_f$. The standard calculation gives the RG equations for 
$m^2=0$\footnote{The overall coefficient of the 2nd order RG equations can always be chosen by a simple redefinition $g_i\rightarrow c g_i$ 
for constant $c$. We have decided to completely remove the factors of $\pi$ which come from the integration over the volume of the 3-sphere.} 
(see e.g. \cite{kolnberger1990critical,Folk:1998yt,Ihrig:2019kfv}), 
\begin{subequations}\label{eqs:RG_flow1}
\begin{align}
&\frac{dg}{d\ell}=-\frac{N_fg^2}{3}\;,\label{eq:RG_flow1g}\\ 
&\frac{d\lambda}{d\ell}=-(N_f+4)\lambda^2+6\lambda g-6 g^2 \;,
\label{eq:RG_flow1_lam}
\end{align}
\end{subequations}
where $g=e^2$, and $\ell$ is the RG flow ``time''\footnote{The ``flow time'' is given by 
$\ell=-log(\mu/\mu_0)$ where $\mu_0$ is the initial mass-dimension cutoff scale, and $\mu<\mu_0$ is the final cutoff scale. }. 
As is commonly done, these RG equations only take into account the marginal couplings, and ignore the infinite set of 
irrelevant couplings that a-priori are not expected to be important. 

The point $\lambda=e^2=0$ is a fixed point. The question is whether this fixed point is reached. If it is, then the transition at $m^2=0$ is 2nd order. 
Naively, since both $g$ and $\lambda$ are marginally irrelevant, it seems this is always true. However, note that it takes infinite flow time 
$\ell$ for $g$ to reach zero, while it is not excluded that $\lambda$ becomes zero in finite flow time. If that happens the last term 
$-6g^2$ in \eqref{eq:RG_flow1_lam} will push $\lambda$ to negative values. When $\lambda$ is negative, however, the would-be irrelevant 
couplings such as $\phi^6$ become important as they stabilize the potential, i.e., they become dangerously irrelevant. If this happens the 
system flows to a Higgs phase at $m^2=0$, while for $m^2>0$ it is in a photon phase, rendering the transition discontinuous.   

This indeed happens for $N_f$ small enough as we explain in the appendix, and in particular for $N_f=1$\footnote{For $N_f=1$, the 
1st order transition also follows from the standard effective potential of Coleman and Weinberg \cite{Coleman:1973jx}.}. On the other hand, if one takes the large 
$N_f$ limit in equations \eqref{eqs:RG_flow1}, implies that the fixed point $e^2=\lambda=0$ is generically reached for $N_f$ sufficiently 
large\footnote{While the large $N_f$ limit seems to indicate that this happens for any initial values, this is in fact not the case. 
The reason is that the decay of $\lambda$ toward zero is faster than the decay of $e^2$, rendering subleading terms at large 
$N_f$ potentially important. A careful analysis is presented in the appendix.}.

Then there should exist a window $1<N_f<N_f^*$ for which the couplings $(e^2, \lambda)$ flow in the infrared 
towards the origin but then generically {\it miss} the mean-field point at zero coupling $(e^2, \lambda) = (0,0)$, instead 
heading off towards large negative coupling $
\lambda$, leading to a 1st order transition. 

However, when $N_f \ge N_f^*$, the dynamics along the RG flow and thus the topology of the solution space is different such that
 there exists an open set of initial data with positive $(e^2, \lambda)$ which are attracted in the far IR towards $(e^2, \lambda) = (0,0)$, 
 rather than generically ``missing'' the origin. Thus one can now arrive at the gapless weak-coupling critical point by tuning only $m^2$, 
 and the transition may be second order, described by a mean-field phase transition with vanishing couplings\footnote{It is interesting 
 to note that the only possible conformally-invariant critical point for a parity-invariant theory with a $\U(1)$ 1-form symmetry -- in 
 this case magnetic flux conservation -- is the free fixed point \cite{Hofman:2018lfz}}. In Appendix \ref{app:RG_analysis} we show that $N_f^*=183$. 

The upshot of this analysis is that for a small number of electric flavors $N_f^e$ (and if magnetic monopoles are heavy) the Higgsing 
transition separating the Coulomb from the Higgs phase is expected to be first-order. By electric-magnetic duality, this also implies that for a small 
number of magnetic flavors $N_f^m$ (and if electric charges are heavy), the confinement transition separating the Coulomb phase from 
a confined phase is expected to be first-order. Thus both of the lines bounding the Coulomb phase in Figure \ref{fig:phase_diagram} are first-order lines. 

We note that in \cite{Kerler:1996cr,Damm:1997vd} it was argued that the introduction of an ad hoc monopole mass term leads to a region of the phase diagram where the confinement-deconfinement transition  is 2nd order. Since this setup is in the same universality class as the Abelian-Higgs model studied here, we conclude that the transition observed in \cite{Kerler:1996cr,Damm:1997vd} is never continuous but is weakly 1st order. In fact in \cite{Damm:1997vd} it was observed that the alleged 2nd order transition is not universal. The likely explanation is the presence of two marginally irrelevant couplings which could keep the system in the vicinity of the non-interacting fixed point until exponentially large volumes are reached, thereby obscuring the 1st order nature of the transition.

\subsection{Generalization of the lattice model, symmetries and anomalies}\label{sec:generalizations}

Here we discuss several generalizations of our lattice model, the symmetries which arise and their 't~Hooft anomalies. 
This section is quite independent from the rest of the paper, and can be skipped at first reading.

The lattice model with one flavor of electric and magnetic matter has very little symmetries, and hence is not very constrained. 
However, it can be generalized in several ways, which introduces more symmetries with 't Hooft anomalies. We will first consider 
the generalization to general charges of the dynamical electric and magnetic matter, and then to introducing flavor multiplets. 
We conclude by formulating a non-abelian gauge theory with a self-dual electric-magnetic symmetry.

\subsubsection{General charge theories and 1-form 't~Hooft anomalies}\label{sec:cent_sym_anom}

We consider coupling matter with charges $q_e$ and $q_m$ larger than 1, i.e., the electric matter fields couple to link phases 
$e^{iq_eA_l}$ and the magnetic matter to $e^{iq_m \tilde A_{\tilde l}}$. The model then has 1-form global symmetries 
$\mathbb Z^{[1]}_{q_e}\times \mathbb Z^{[1]}_{q_m}$, as is clear from the invariance of the action under the transformation
\begin{align}\label{eq:el_1form_sym}
&A^e_l\rightarrow A^e_{l}+\frac{2\pi k_l}{q_e}\;,\\
&A^m_{\tilde l}\rightarrow \tilde A^m_{\tilde l}+\frac{2\pi \tilde k_{\tilde l}}{q_m}\;,
\end{align}
where $k_l$ and $\tilde k_{\tilde l}$ are such that
\be
(dk)_p=0 \quad \forall \; p\in \Lambda, 
\qquad \mbox{and} \qquad (d\tilde k)_{\tilde p} = 0
\quad \forall  \; \tilde p\in \tilde\Lambda\;.
\ee
We can also consider introducing background gauge fields for the 1-form symmetries. As we now show, there is an obstruction to putting 
background gauge fields for both 1-form symmetries simultaneously, which is a manifestation of the 't~Hooft anomaly. Indeed, if we want 
to put background fields for the electric center symmetry, we must promote the shifts \eqref{eq:el_1form_sym} to be valid for any $k_l$, 
not necessarily only those obeying the constraint $dk=0$. We do this by introducing a background 2-form gauge field  (i.e., living on plaquettes) 
$B_p^e=\frac{2\pi b^e_p}{q_e}, b_e\in\mathbb Z$, and replacing in the gauge action\footnote{Notice that despite $b_e$ taking all 
integer values, it should be thought of as a $\mathbb Z_{q^e}$ gauge field, because shifts of $b_e\rightarrow b_e+q_e$ 
can be absorbed by a corresponding shift of the dynamical Villain variables $n_p\rightarrow n_p-1$.}
\be
(dA^e)_{p}+2\pi n_p\rightarrow (dA^e)_{p}+2\pi n_p+\frac{2\pi b^e_p}{q_e}\;.
\ee
The 1-form electric center symmetry is now a gauge symmetry 
\begin{align}\label{eq:el_1form_g_sym}
&A^e_l\rightarrow A^e_{l}+\frac{2\pi k_l}{q_e}\;,\\
&b^e_p\rightarrow b^e_p-(dk)_p\;.
\end{align}

However, note that now $B_p^e$ is supposed to be a $\mathbb Z_{q_e}$-valued gauge field, which means that setting $b^e_p$ to be an 
integer multiple of $q_e$, i.e., $b^e_p=m_p q_e,\; m_p\in \mathbb Z$, should be the same as not putting a background at all. Indeed 
for a gauge-kinetic term, we can always shift $n_p\rightarrow n_p-m_p$ to absorb such a field, but must recall that $n_p$ also 
appears in the Lagrange multiplier term involving the magnetic field $\tilde A^m_{\tilde l}$. Hence we also must replace
\be
i\sum_p(d\tilde A^m)_{\star p}n_p\rightarrow  i\sum_p(d\tilde A^m)_{\star p}\left(n_p+\frac{b^e_p}{q_e}\right)\;.
\ee
So far so good. We have managed to put a consistent background gauge field for a $1$-form electric center symmetry. 
But now we find that in order to promote the magnetic center symmetry to a gauge symmetry, we must replace
\be
(d\tilde A^m)_{\tilde p}\rightarrow \left((d\tilde A^m)_{\tilde p}+\frac{2\pi \tilde b^m_{\tilde p}}{q_m}\right)\;, \qquad \tilde b^m_{\tilde p}\in \mathbb Z\;.
\ee
In particular the Lagrange multiplier term in the action becomes
\be
i\sum_p(d\tilde A^m)_{\star p}\left(n_p+\frac{b^e_p}{q_e}\right)\rightarrow i\sum_p\left((d\tilde A^m)_{\star p}+
\frac{2\pi \tilde b^m_{\star p}}{q_m}\right)\left(n_p+\frac{b^e_p}{q_e}\right) \; .
\ee
But in the above, the shift $\tilde b^m_{\tilde p}$ by arbitrary integer multiples of $q_m$ is not a symmetry, as such shifts 
will get a contribution from the cross term between $\tilde b^m$ and $b^e$. Indeed we find that under 
$\tilde b_{\tilde p}^m\rightarrow \tilde b_{\tilde p}^m+r_{\tilde p} q_m, r_{\tilde p}\in \mathbb Z$, the action changes by
\be
i 2\pi  \frac{r_{\star p}}{q_e} b^e_{p}\;.
\ee
Can this non-invariance be fixed by a local counter-term? Indeed such a term should be linear in $b^e$ and $\tilde b^m$ to reproduce the transformation above. Moreover, it should be invariant under the shift of $b_p^e$ by integer multiples of $q_e$. The only such counter-term is given by
\be
 S_{counter}=-i\frac{2\pi P}{q_e}\sum_p\tilde b^m_{\star p} b^e_{p}\;, \quad P=0,1,2,\dots, q_e\;.
\ee
Now the shift $b_{\tilde p}^m \rightarrow \tilde b_{\tilde p}^m+ r_{\tilde p}q_m$ changes the counter-term as follows
\be
\Delta S_{counter}= -i\frac{2\pi P q_m}{q_e}\sum_{p}r_{\star p}b^e_{p}\;,
\ee
which restricts $P$ to be such that
\be
P q_m=1 \bmod q_e\;.
\ee
If such a $P$ can be found then there is no mixed anomaly between the two 1-form symmetries. In particular the above 
condition translates to $\text{GCD}(q_m,q_e)=1$. So we found that there is a mixed 't~Hooft anomaly if $\text{GCD}(q_m,q_e)\ne 1$.

\subsubsection{General numbers of electric and magnetic flavors}

We now consider $SU(N_f^e)$ and $SU(N_f^m)$ matter multiplets, which can be either fermionic or bosonic. 
We couple these to the electric and magnetic gauge fields respectively. This is done in the standard way by with a gauge invariant hopping term
\be
\sum_{x,\mu} \,  \sum_{I=1}^{N_f^e}  \, \phi^{I\,*}_x \; e^{\, i \, A^e_{x,\mu}} \; \phi^I_{x+\hat \mu} \; + \; c.c.\;,
\ee
where $\phi_x$ is the a matter field on the lattice, coupling to an electric gauge field $A^e$. 
The index $I$ is an $SU(N_f^e)$ flavor index. A similar expression can be written for the magnetic multiplet,
\be\label{eq:magnetic_coupling}
\sum_{\tilde x,\mu} \, \sum_{I=1}^{N_f^m} \, {\tilde\phi}^{I\,*}_{\tilde x} \; e^{\, i \, \tilde A^m_{\tilde x, \mu} } \; \tilde\phi^I_{\tilde x + \hat \mu} \; + \; c.c.\;.
\ee 
Note that the global symmetry is not $SU(N_f^e)\times SU(N_f^e)$ but $PSU(N_f^e)\times PSU(N_f^e)$\footnote{$PSU(N)\equiv SU(N)/\mathbb Z_N$.}, 
because the transformation by the center element which takes $\phi_x^I \rightarrow \exp\left(\frac{2\pi i}{N_f^e}\right)\phi_x^I$ is part of the 
$\U(1)$ electric gauge transformation, and similarly for the transformation with the center of $SU(N_f^m)$. 

Now let us couple the symmetry to a background gauge field. We promote  
\be
\sum_{x,\mu} \,  \sum_{I=1}^{N_f^e}  \, \phi^{I\,*}_x \; e^{\, i \, A^e_{x,\mu}} \; \phi^I_{x+\hat \mu} \; 
\rightarrow \; \sum_{x,\mu} \,  \sum_{I,J=1}^{N_f^e}  \, \phi^{I\,*}_x \; U^{e, IJ}_{x,\mu} \; \phi^J_{x+\hat \mu} 
\ee
where $U^{e}_{x,\mu}$ is an $SU(N^e_f)$ matrix. However, recall that the symmetry we wish to gauge is really 
$PSU(N_f^e)$, while our link matrices $U^{e}_{x,\mu}$ are $SU(N_f^e)$. To do this we must make sure that there exists a gauge symmetry 
$U^{e}_{x,\mu} \rightarrow \exp\left({i\frac{2\pi s_{x,\mu}}{N_f^e}}\right) U^{e}_{x,\mu}$, with $s_{x,\mu}=1,\dots,N_f^e-1$, 
which effectively removes the center of $SU(N_f)$ and 
renders the gauge field a $PSU(N_f^e)$ gauge field. This can be accomplished by shifting 
$A_{x,\mu}^e\rightarrow A_{x,\mu}^e-\frac{2\pi s_{x,\mu}}{N_f^e}$, 
which is not a symmetry of the kinetic term however. To turn it into a symmetry we introduce a background 2-form 
$\mathbb Z_{N_f^e}$ gauge field $b_p^e$ like before and replace the kinetic term and the Lagrange multiplier term by
\begin{align}
&\Big((dA^e)_{p}+2\pi n_p\Big)^2\rightarrow \left((dA^e)_{p}+2\pi n_p+\frac{2\pi b^e_p}{N_f^e}\right)^2\;,\\
&i(d\tilde A^m)_{\star p}\left(n_p+\frac{b^e_p}{q_e}\right)\rightarrow i(d\tilde A^m)_{\star p}\left(n_p+\frac{b^e_p}{N_f^e}\right)\;.
\end{align}
Note that the gauge invariant flux is now $(dA^e)_{p}+2\pi n_p+\frac{2\pi b^e_p}{N_f^e}$, and can be 
fractional\footnote{We can additionally demand that $(db^e)_c=0\bmod 2\pi$, i.e., that $b^e_p$ is a representative of 
$H^{2}(M,\mathbb Z_{N_f^e})$, where $M$ is the underlying space-time manifold. This will not change any arguments 
below. Then $b^e_p$ is a representative of the well-known characteristic class 
$w_2\in H^{2}(M,\mathbb Z_{N_f^e})$ of the $PSU(N_f^e)$ principal bundle.}. 

Now the background gauge field $U^{e}_{x,\mu} \in SU(N_f^e)$ has its center promoted to a gauge symmetry, and hence represents a $PSU(N_f^e)$ gauge background. As part of this background we had to introduce 2-form gauge fields which are meaningful mod $N_f^e$, i.e.,
$b^e_p=0,1,\dots, N_f^e-1 \bmod N_f^e$. The remaining steps closely follow the discussion of the 't Hooft anomaly with 
1-form symmetries. We spell them out for convenience\footnote{A way to understand the similarity is to introduce $SU(N_f^e)$ gauge fields without 
introducing the $b^e$ field. In the presence of $SU(N_f^e)$ gauge fields only, the setup has a global bonus 1-form $\mathbb Z_{N_f^e}^{[1]}$ 
symmetry, and the anomaly analysis largely reduces to the analysis of the global 1-form symmetries we discussed previously. 
This bonus 1-form symmetry is precisely there because by considering only $SU(N_f^e)$ background fields, we effectively 
ignore $PSU(N_f^e)$ gauge bundles which cannot be lifted to $SU(N_f^e)$ gauge bundles. Then we can think of placing 
background gauge fields $b^e$ for this bonus 1-form symmetry, which is the same as introducing the 
$w_2\in H^2(M,\mathbb Z_{N_f^e})$ classes we discussed previously. }. 

We first repeat the same for $PSU(N_f^m)$, and replace the expression \eqref{eq:magnetic_coupling} with
\be\label{eq:magnetic_coupling_gauged}  
\sum_{\tilde x,\mu} \, \sum_{I,J=1}^{N_f^m} \, {\tilde\phi}^{I\,*}_{\tilde x} \; U^{m,IJ}_{\tilde x, \mu}  \; \tilde\phi^J_{\tilde x + \hat \mu} \; + \; c.c.\;.
\ee
Again we have to promote the shift of an $SU(N_f^m)$ by a center to a gauge symmetry, 
$U^{m}_{\tilde x, \mu} \rightarrow \exp\left({\frac{2\pi  i s_{\tilde x,\mu}}{N_f^m}}\right)U^{m}_{\tilde x, \mu}$ 
must be a symmetry for any $s_{\tilde x,\mu} =1,2,\dots N_f-1$ and for  
any link $(\tilde x,\mu)$. Now we have a problem similar to the one we encountered when we spoke about center symmetries. To gauge 
the full $PSU(N_f^m)$ we must replace $(d\tilde A^m)_{\tilde p}\rightarrow  (d\tilde A^m)_{\tilde p}+\frac{2\pi \tilde b_{\tilde p}}{N_f^m}$, 
where $b_{\tilde p}\in \mathbb Z$ serves as a $\mathbb Z_{N_f^m}$ 2-form gauge field, i.e., we replace
\be
i\sum_p(d\tilde A^m)_{\star p}\left(n_p+\frac{b^e_p}{N_f^e}\right)
\rightarrow i\sum_p\left((d\tilde A^m)_{\star p}+\frac{2\pi \tilde b^m_{\star p}}{N_f^m}\right)\left(n_p+\frac{b^e_p}{N_f^e}\right)\;.
\ee
Here we have the same problem as in the case of higher charged matter, with the similar conclusion that an `t Hooft 
anomaly between $PSU(N_f^e)$ and $PSU(N_f^m)$ exists if $\text{GCD}(N_f^e,N_f^m)\ne 1$. 

\subsubsection{The self-dual non-abelian gauge theory}

This will be our last generalization, and it is a small change on the multi-flavor theory we just discussed. The idea is to now promote the $SU(N_f^e)$ and $SU(N_f^m)$ link field $U_l^e$ and $U_l^{m}$ gauge background into a dynamical one. This leaves the theory as having two 1-form symmetries $\mathbb Z^{[1]}_{ N_f^e}\times \mathbb Z^{[1]}_{ N_f^m}$. If $\text{GCD}(N_f^e,N_f^m)\ne 1$, the two 1-form symmetries have mixed `t Hooft anomalies. This then is a model which is a fully fledged non-abelian gauge theory with gauge group $SU(N_f^e)\times SU(N_f^m)$, and a self-dual symmetry exchanging the two non-abelian gauge groups $SU(N_f^e)$ and $SU(N_f^m)$. This theory is extremely interesting, as it is a fully non-abelian theory, with electric-magnetic duality. However, all these theories have a complex-action problem, and cannot be simulated in the form we so far used, and so-called worldline representations of either electric or magnetic matter must be used. If the matter is charged under a dynamical gauge group, it is not clear that worldline representation will be useful in solving the complex-action problem.

\subsection{The phase diagram of self-dual scalar QED}\label{sec:phase_diagram}

\begin{figure}[!tbp] 
   \centering
   \includegraphics[width=4.5in]{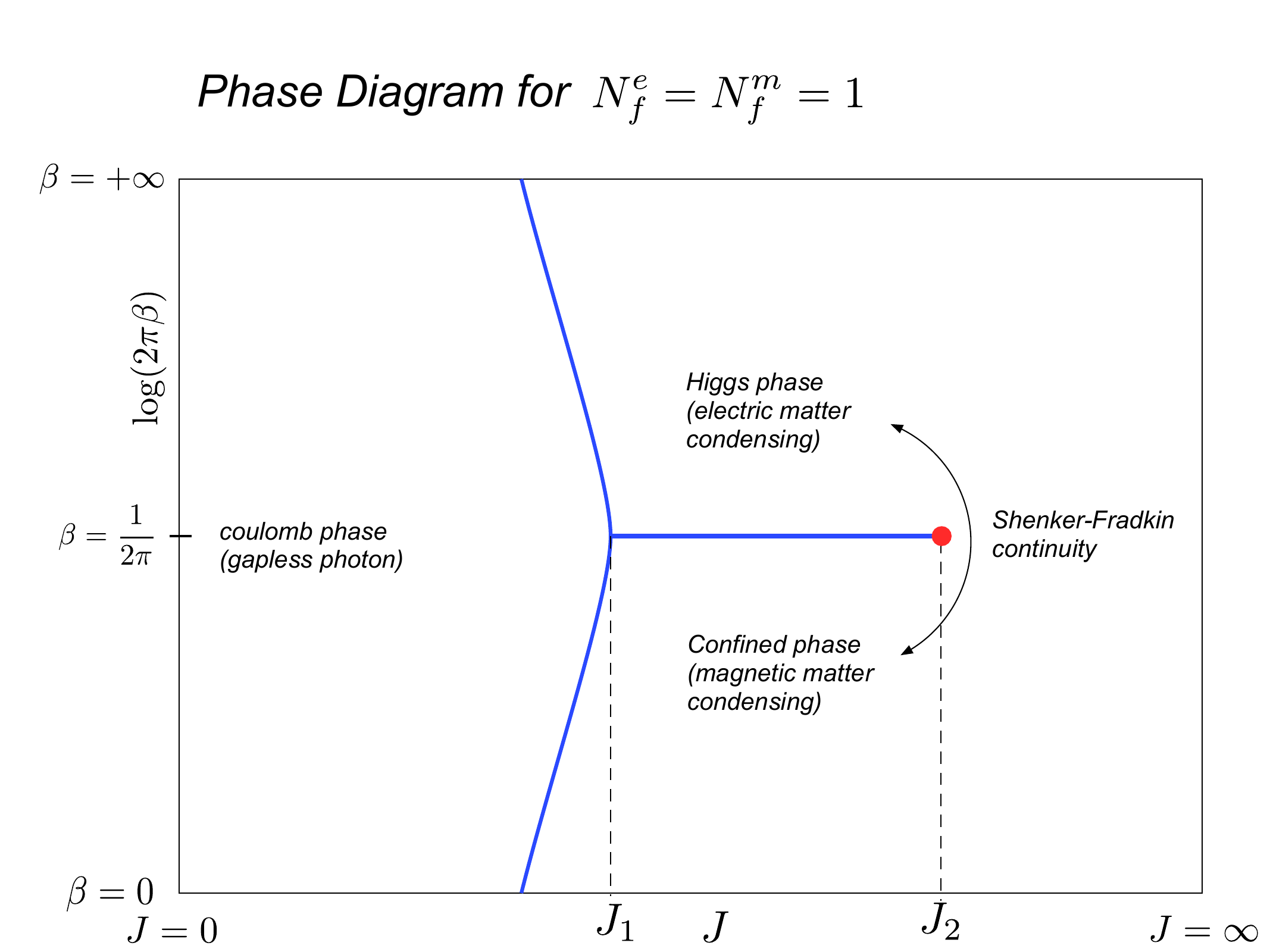} 
   \caption{A cartoon of the phase diagram in the $J$, $\log(2\pi\beta)$ plane. }
   \label{fig:phase_diagram}
\end{figure}

Here we collate all of the information we discussed so far to come up with a phase diagram for the self-dual lattice theories. We focus primarily on 
$N_f^e=N_f^m=1$ with only bosonic matter, whose phase diagram is sketched in Fig.~\ref{fig:phase_diagram}. We plot the diagram as 
a function of $J=J^e=J^m$ and $\log(2\pi\beta)$, because under self-duality $\log(2\pi\beta)\rightarrow -\log(2\pi\beta)$, such 
that the diagram has to have a symmetry around the line $\log(2\pi\beta)=0$, which is the self-dual line. 

On the one hand we said that as $J^e=J^m=J\rightarrow\infty$ we have a gapped phase, with no phase transition as the coupling $\beta$ is varied. 
This is illustrated on the far right of the diagram. The phase $J=0$ is a free photon phase, which is the far left of the diagram. 
The limits $\beta\rightarrow \infty$ and $\beta\rightarrow 0$ are treatable perturbatively in the Abelian-Higgs model, 
and exhibit a 1st order transition for $N_f=1$. 

Now let us consider the self-dual line $\beta=1/(2\pi)$ and change $J$ from zero to infinity. Since the theory will be trivially gapped 
for $J$ sufficiently large, we expect a phase transition. In fact dialing $J$ to higher values is driving the system towards preferring a 
condensation of matter. But since condensing electric matter (the Higgs phase) confines magnetic matter, and condensing magnetic 
matter (the confined phase) confines electric matter, a tension is expected between the condensation of electric and magnetic matter 
so that a phase coexistence line should emerge at some value $J=J_1$. Since the electric and magnetic condensed phases coexist, 
this regime spontaneously breaks the self-dual symmetry. This is depicted by the horizontal blue line segment in Fig.~\ref{fig:phase_diagram}.

On the other hand if we crank up $J$ sufficiently high, we know that eventually the phase coexistence will disappear, 
as we discussed in Sec.~\ref{sec:limits}, and the coexistence phase should disappear at some value $J = J_2>J_1$. 
The disappearance of the 1st order line is a critical point. Since the critical point restores the discrete self-dual symmetry, we expect this to be in the 4d Ising universality class, which is a Gaussian fixed point. Indeed our numerical results will agree with this.

A more interesting question is what happens at the transition between the free-photon phase and the self-dual broken phase, i.e., 
the leftmost point of the horizontal segment in Fig.~\ref{fig:phase_diagram}. This is the point where the 1st order transitions which 
were computable in the $\beta=0$ and $\beta=\infty$ limits, meet at the self-dual point. As we will see, numerical computations
 reveal this point to be a triple point, i.e., a coexistence point of three phases.

\begin{figure}[!tbp] 
\centering
\includegraphics[width=4.5in]{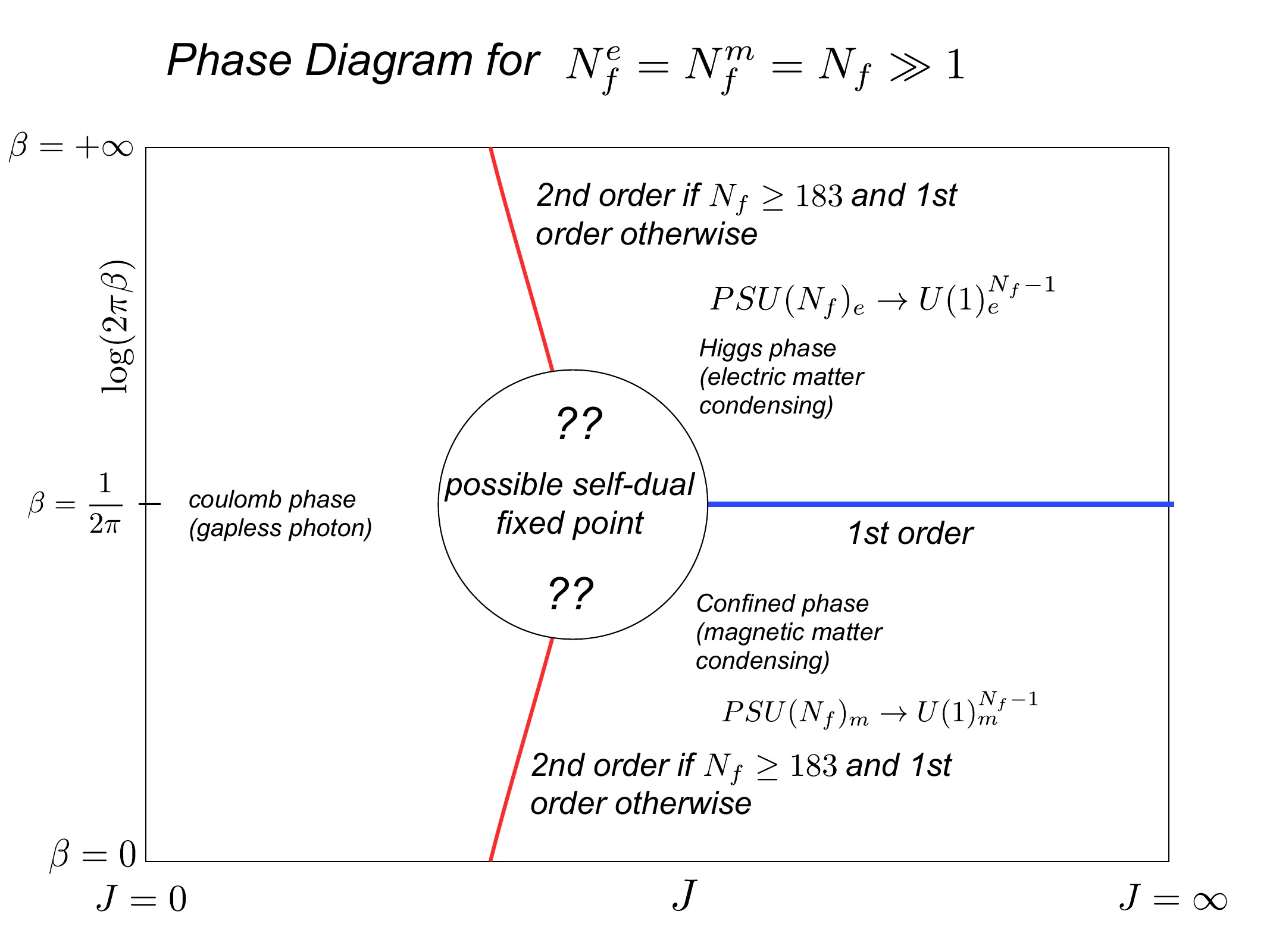} 
\caption{The conjectured phase diagram for sufficiently large $N_f^e=N_f^m=N_f$. The limits $\beta\sim 0$ or $\beta\sim +\infty$ 
can have a 2nd order phase transition if $N_f\ge 183$, as discussed in Sec.~\ref{sec:limits}.}
\label{fig:phase_diag_largeNf}
\end{figure}

Let us quickly discuss the phase diagram for other generalizations. We first consider $N_f^e=N_f^m=1$, theory with 
charges $q^m=q^e=q>1$. As we discussed, this theory has a mixed anomaly between $\mathbb Z_{q}^{[1]}$-electric and
 $\mathbb Z_{q}^{[1]}$-magnetic 1-form symmetries. These anomalies are matched by either a photon phase or by spontaneous 
 breaking of electric and/or magnetic 1-form symmetries. If the $\mathbb Z_q^{[1]}$-electric symmetry is spontaneously broken, 
 we call that the Higgs phase, while the confined phase spontaneously breaks $\mathbb Z_{q}^{[1]}$-magnetic. However, 
now these two phases cannot be continuously connected, because they break different symmetries, so instead of the 
coexistence line ending, we expect that it continues all the way to $J=\infty$. 

A similar picture is expected for $N_f^e=N_f^m=N_f>1$ theories, except that now the Higgs and confined phases are respectively 
breaking $PSU(N_f)$-electric and $PSU(N_f)$-magnetic global symmetries, which will be gapless. Another point we should make 
is that for sufficiently large $N_f$, we saw that the regimes $\beta\rightarrow 0$ and $\beta\rightarrow \infty$ have 2nd order phase 
transitions, and so it is possible that this regime continues all the way to the self-dual point. This would indicate the existence of a new, 
self-dual fixed point. This is sketched in Fig.~\ref{fig:phase_diag_largeNf}. A similar phenomenon is known to occur for $\mathbb{Z}_2$ 
gauge theory in 3d, where it is believed that two continuous Ising lines meet at a novel self-dual critical point \cite{Somoza:2020jkq}.

\section{Numerical simulation}\label{sec:numerics}

\subsection{Switching to a dual worldline formulation}
\label{subsec:worldlines}

We now want to study the minimal interacting self-dual theory with $N_f^e=N_f^m=1$ numerically and analyze the phase diagram 
sketched in Fig.~\ref{fig:phase_diagram}. However, self-dual $\U(1)$ gauge theory coupled to electric and magnetic matter as introduced 
in Sec.~\ref{sec:coupling_to_matter} is not yet suitable for a numerical 
simulation, since the gauge field Boltzmann factor (\ref{Zfull}) has a complex phase and does not give rise to a real and positive
weight that can be used in a Monte Carlo simulation. In this subsection we now show that this complex action 
problem can be overcome by switching to a worldline formulation for the magnetic matter.

In order to prepare the Boltzmann factor (\ref{boltzmann_both}) for the worldline formulation we rewrite the second exponent in 
(\ref{boltzmann_both}) by switching to the dual lattice using the identity\footnote{We remark that the step of switching to the dual 
Villain variables is not essential but simplifies the discussion here and also the actual computer code used in the simulations discussed in 
Subsections 3.3 and 3.4.}
\begin{equation}
(dn)_{x,\mu\nu\rho} \; = \; - \sum_{\sigma} \epsilon_{\mu\nu\rho\sigma} \, (\, \partial \, \widetilde{n}\,)_{\tilde{x}-\hat{\sigma},\sigma} \; ,
\label{derivrelation}
\end{equation}
which is straightforward to check (see the appendix of \cite{Anosova:2022cjm}). Thus the gauge field Boltzmann factor assumes the form 
\begin{equation}
B_\beta[A^e,A^m]  \; = \; \sum_{\{ n \}} \; e^{ \, - \beta \, S_g[A^e\!,\,n] }
\; \prod_{\tilde{x},\mu} e^{\, - i  \widetilde{A}^m_{\tilde{x},\mu} (\partial \widetilde{n})_{\tilde{x},\mu}}  \, ,
\label{boltzmann_both_dual}
\end{equation}
where we have converted the sum in the second exponent of the Boltzmann factor into a product over all links of the dual lattice.

The second step is to use the well known worldline representation for $\U(1)$ gauge field theories
(see, e.g., \cite{Mercado:2013yta,Mercado:2013ola}). It is straightforward to convert this worldline representation to the dual lattice 
where the magnetic matter partition sum (\ref{Zmagnetic}) is defined. The worldline representation then reads (compare the appendix
of \cite{Sulejmanpasic:2019ytl} for the notation used here)
\begin{equation}
\widetilde{Z}[\widetilde{A}^m, J_m\big] \; = \; 
\sum_{\{ \widetilde{k} \}} \, \left[ \prod_{\tilde{x},\mu} I_{\widetilde{k}_{\tilde{x},\mu}} (J_m) \right] \left[
\prod_{\tilde{x}} \delta \Big( \big(\partial \widetilde{k}\,\big)_{\tilde{x}} \Big) \right] \left[
\prod_{\tilde{x},\mu} e^{\, i \, \widetilde{A}^m_{\tilde{x},\mu} \, \widetilde{k}_{\tilde{x},\mu} } \right] \;  ,
\label{Zm_worldline}
\end{equation}
where $I_n(x)$ denotes the modified Bessel functions. 
The partition function is a sum over configurations of the dual flux variables $\widetilde{k}_{\tilde{x},\mu} \in \mathds{Z}$ assigned to the 
links $(\tilde{x},\mu)$ of the dual lattice, where 
\begin{equation}
\sum_{\{ \widetilde{k} \}} \; \equiv \; \prod_{\tilde{x},\mu} \; \sum_{\widetilde{k}_{\tilde{x},\mu} \in \mathds{Z}} \; .
\label{fluxsum}
\end{equation}
The flux variables are subject to vanishing divergence constraints 
\begin{equation}
\big(\partial \widetilde{k}\,\big)_{\tilde{x}} \; \equiv \; \sum_{\mu = 1}^d 
\Big[ \widetilde{k}_{\tilde x,\mu} - \widetilde{k}_{\tilde x-\hat \mu,\mu} \Big] \; = \; 0 \; \; \; \; \forall \; \tilde x \; ,
\label{divergence0}
\end{equation}
which in (\ref{Zm_worldline}) are implemented with the product of Kronecker deltas. These constraints  enforce flux conservation 
at each site $\tilde x$ of the dual lattice, such that the $\widetilde{k}_{\tilde{x},\mu}$ form closed loops
of flux on the dual lattice. At every link $(\tilde{x},\mu)$ of the dual lattice the dual magnetic gauge field $\widetilde{A}^m_{\tilde{x},\mu}$
couples in the form  $e^{\, i \, \widetilde{A}^m_{\tilde{x},\mu} \, \widetilde{k}_{\tilde{x},\mu} }$ which gives rise to the second product in 
(\ref{Zm_worldline}). The configurations of the dual flux variables $\widetilde{k}_{\tilde{x},\mu}$ come with real and
 positive weight factors given by the Bessel functions. 

With the gauge field Boltzmann factor in the form (\ref{boltzmann_both_dual}) and the dependence of the partition sum 
$\widetilde{Z}[\widetilde{A}^m, J_m\big]$ on the dual magnetic gauge field $\widetilde{A}^m_{\tilde{x},\mu}$ given by the 
last factor in (\ref{Zm_worldline}) we can now completely integrate out the dual magnetic gauge field. The corresponding 
integral reads (compare (\ref{Zfull}) and use $\int \! D[{A}^m] = \int \! D[\widetilde{A}^m]$ ), 
\begin{eqnarray}
\int \!\! D[\widetilde{A}^m] \!
\left[\prod_{\tilde{x},\mu} e^{\, - i  \widetilde{A}^m_{\tilde{x},\mu} (\partial n)_{\tilde{x},\mu}} \right] \!\!
\left[\prod_{\tilde{x},\mu} \, e^{\, i \, \widetilde{A}^m_{\tilde{x},\mu} \, \widetilde{k}_{\tilde{x},\mu} } \right]  & = & 
\prod_{\tilde{x},\mu} \int_{-\pi}^\pi \!\! \frac{ d \widetilde{A}^m_{\tilde{x},\mu} }{2\pi} 
e^{\,  i  \widetilde{A}^m_{\tilde{x},\mu} \big[ \widetilde{k}_{\tilde{x},\mu}  -  (\partial \widetilde{n})_{\tilde{x},\mu} \big] }
\nonumber \\ 
 & = & \prod_{\tilde{x},\mu} \delta \Big( \widetilde{k}_{\tilde{x},\mu}  -  ( \partial \widetilde{n})_{\tilde{x},\mu} \Big) \; .
\label{kidentity}
\end{eqnarray}
Integrating out the dual magnetic gauge fields has generated link-based constraints that completely determine the flux variables
as 
\begin{equation}
\widetilde{k}_{\tilde{x},\mu} \; =  \; ( \partial \widetilde{n})_{\tilde{x},\mu}
\qquad \forall (\tilde x, \mu) \; .
\label{ksolution}
\end{equation}
Note that the configurations (\ref{ksolution}) also obey the vanishing divergence constraints $\partial \widetilde{k} = 0$ from 
(\ref{divergence0}), due to $\partial^2 = 0$ (see the appendix of \cite{Anosova:2022cjm}). 
 
Thus we may summarize the final form of self-dual scalar lattice QED with a worldline representation for the magnetic matter: 
\begin{equation}
Z(\beta, J_e, J_m) \; = \, \int \!\! D[A^e]  \sum_{\{ n \}}  \int \!\! D[\phi^e] \; 
e^{ \, - \, \beta \, S_g[A^e\!,\,n] \; + \; J_e S_e [\phi^e,\, A^e ]} \; 
\left[ \prod_{\tilde{x},\mu} I_{(\partial \widetilde{n})_{\tilde x,\mu}} (J_m) \right] \; .
\label{Zfull_worldline1}
\end{equation}
Obviously all weight factors in (\ref{Zfull_worldline1}) are real and positive, such that this form now is accessible to numerical 
Monte Carlo simulations. Note that here the Villain variables are not subject to any constraints, which in some aspects makes a 
numerical simulation of  (\ref{Zfull_worldline1}) simpler than the simulation of the pure gauge theory 
(\ref{eq:part_funct}), where configurations of the 
Villain variables need to obey the closedness constraint (\ref{closedness}). 

It is interesting to consider the limit $J_m \rightarrow 0$. Using the fact that for the Bessel functions 
$\lim_{x \rightarrow 0} I_0(x) = 1$ and $\lim_{x \rightarrow 0} I_n(x) = 0 \; \; \forall \; n \neq 0$, one finds that in (\ref{Zfull_worldline1})
only those configurations of the Villain variables survive where the dual Villain variables obey
\begin{equation}
(\partial \widetilde{n})_{\tilde x,\mu} \; = \; 0 \; \; \forall \; (\tilde x,\mu) \qquad \Longleftrightarrow \qquad 
(dn)_{x,\mu\nu\rho} \; = \; 0 \; \; \forall \; (x,\mu\nu\rho) \; , 
\end{equation}
where in the second form we used (\ref{derivrelation}) to identify this constraint as the closedness condition for the 
Villain variables on the original lattice. Thus we find
\begin{equation}
\lim_{J_m \rightarrow 0} \, Z(\beta, J_e, J_m) \; = \, \int \!\! D[A^e]  \sum_{\{ n \}} 
e^{ \, - \beta \, S_g[A^e\!,\,n] } \; Z[A^e,J_e] \;
\left[\prod_{x} \prod_{\mu<\nu<\rho} \delta\big((dn)_{x,\mu\nu\rho}\big) \right]\; .
\label{Zfull_worldline2}
\end{equation}
Although not self-dual, this is an interesting theory in its own right, as it describes $\U(1)$ lattice gauge fields coupled to 
electric matter without magnetic monopoles that appear in the usual lattice discretization of this system. Finally we remark
that a second limit $J_e \rightarrow 0$ reduces the partition sum to our pure gauge theory partition sum 
(\ref{eq:part_funct}) without monopoles. 

We conclude the discussion of the worldline form by expressing the expectation value 
$\left\langle \widetilde{s}_m \right\rangle_{\beta,J_e,J_m}$ that
appears in the duality relation (\ref{duality_phi2}) in terms of the worldline variables. The expectation value is obtained from
a derivative of $\ln Z$ with respect to $J_m$, and this derivative can of course also be applied to $Z(\beta, J_e, J_m)$ 
in the form (\ref{Zfull_worldline1}). A few lines of algebra give 
\begin{equation}
\left\langle \widetilde{s}_m \right\rangle_{\beta,J_e,J_m} \; = \; 
\frac{1}{4 V} \frac{\partial }{\partial J^m} \, \ln Z(\beta, J_e, J_m) \; = \; 
\frac{1}{4 V} \left\langle \sum_{\tilde x, \mu} \frac{I_{(\partial \widetilde{n})_{\tilde x,\mu}}^{\, \prime} \!(J_m)}
{ I_{(\partial \widetilde{n})_{\tilde x,\mu}} \! (J_m) }
\right \rangle_{\!\! \beta,J_e,J_m}  \; .
\end{equation}

\subsection{The self-dual point revisited}
\label{sec:selfdual}

To prepare for the numerical simulations presented in the next two subsections we here discuss suitable 
observables at the self-dual point of the inverse gauge coupling, i.e., at 
\begin{equation} 
\beta \; = \; \widetilde{\beta} \; = \; \beta^* \; \equiv \; \frac{1}{2\pi} \; .
\label{couplings_sd1}
\end{equation}
Furthermore we may set $J_e$ and $J_m$ equal to the same value $J$ such that 
\begin{equation}
J_e \; = \; J_m \; = \; \widetilde{J}_e \; = \;  \widetilde{J}_m \; = \; J \; .
\label{couplings_sd2}
\end{equation}
In other words, the theory has only one remaining parameter, i.e., the coupling $J$. 

With this choice for the couplings the self-duality relation (\ref{sumruleF2_qed}) simplifies to, 
\begin{equation}
\beta^* \, \langle F^2 \rangle_{\beta^*\!, \, J} \; + \; \beta^* \, 
\langle F^2 \rangle_{\beta^*\!, \, J} \;  = \; 1\; ,
\end{equation}
which implies that $\langle F^2 \rangle_{\beta^*\!, \,J}$ is constant,
\begin{equation}
\langle F^2 \rangle_{\beta^*\!, \, J}  \;  = \; \pi \; \; \; \; \forall \; J \; .
\label{F2constant}
\end{equation}
We remark, that the self-duality relation (\ref{sumrulechi_qed}) for the second moment of $F^2$ does not constrain
the susceptibiliy for the self-dual couplings (\ref{couplings_sd1}), (\ref{couplings_sd2}).

The self-duality relation (\ref{duality_phi2}) that links the electric and the magnetic action densities, for the self-dual 
couplings (\ref{couplings_sd1}), (\ref{couplings_sd2}) assumes the form
\begin{equation}
\left\langle s_e \right\rangle_{\beta^*\!, \, J} \; = \; 
 \left\langle \widetilde{s}_m \right\rangle_{\beta^*\!, \, J} \; \; \; \; \forall \; J \; .
\label{duality_phi2_sd}
\end{equation}
An interesting question, already touched upon in the previous subsections, is wether self-duality can be broken spontaneously as a function of
$J$. Such a symmetry breaking should become manifest in a violation of the two relations (\ref{F2constant}) and 
(\ref{duality_phi2_sd}).

For the further analysis we introduce the two order parameters
\begin{equation}
M_g \; \equiv \;  | F^2 \; - \; \pi | \qquad \mbox{and} \qquad M_m \; \equiv \; | s_e \, - \, s_g | \; ,
\label{orderparam}
\end{equation}
which are normalized such that 
\begin{equation}
\langle M_g \rangle_{\beta^*\!, \, J}  \; \neq \; 0 \qquad \mbox{and} \qquad \langle M_m \rangle_{\beta^*\!, \, J}  \; \neq \; 0 \; ,
\end{equation}
signal the breaking of self-duality. The absolute value in the definitions (\ref{orderparam}) was introduced to 
allow for a non-zero expectation value also on a finite lattice. We will also analyze the corresponding susceptibilities
\begin{equation}
\chi_g \; \equiv \; V \, \left\langle \Big( M_g - \langle M_g \rangle_{\beta^*\!, \, J} \Big)^2 \right\rangle_{\beta^*\!, \, J} 
\qquad \mbox{and} \qquad
\chi_m \; \equiv \; V \, \left\langle \Big( M_m - \langle M_m \rangle_{\beta^*\!, \, J} \Big)^2 \right\rangle_{\beta^*\!, \, J}  \; ,
\end{equation}
as well as the Binder cumulants
\begin{equation}
U_g \; \equiv \; 1\; - \; \frac{\big\langle (M_g)^4 \big\rangle_{\beta^*\!, \, J}}{3 \, \big\langle (M_g)^2 \big\rangle_{\beta^*\!, \, J}^{ 2}}
\qquad \mbox{and} \qquad
U_m \; \equiv \; 1 \; - \; \frac{\big\langle (M_m)^4 \big\rangle_{\beta^*\!, \, J}}{3 \, \big\langle (M_m)^2 \big\rangle_{\beta^*\!, \, J}^{ 2}} \; .
\end{equation}
 
\subsection{Setup of the computation and general results }\label{sec:setup}

In this subsection we present our results for the simulation of the full theory at the self-dual point as discussed in 
Subsection \ref{sec:selfdual}, i.e., at $\beta = \beta^* \equiv 1/2\pi$ with $J_e = J_m = J$.  In a second study 
we keep $J$ fixed and vary $\beta$ in the vicinity of $\beta^*$ for studying the nature of the transition when crossing 
the critical line of the phase diagram shown in Fig.~\ref{fig:phase_diagram}.

These simulations are based on the partition sum in the
worldline form (\ref{Zfull_worldline1}), which uses the electric gauge fields $A^n_\mu(x) \in [-\pi, \pi)$ and the Villain variables 
$n_{x,\mu\nu} \in \mathds{Z}$ for the gauge field degrees of freedom, and $\phi^e_x = e^{i\varphi_x}, \varphi_x \in [-\pi, \pi)$ for
the electric matter. There are no constraints left for these variables and they can be updated efficiently using local Metropolis updates,
which we organize in sweeps, i.e., one Metropolis update of all degrees of freedom. For equilibration we use $10^6$ sweeps followed
by $10^5$ measurements of our observables separated by 20 sweeps for decorrelation. For the finite size scaling analysis 
of the critical exponents the number of measurements is increased to $10^6$. The error bars we show are statistical
errors determined with the jackknife method combined with a blocking analysis.

\begin{figure}[p] 
   \hspace*{0mm}
   \includegraphics[width=80mm]{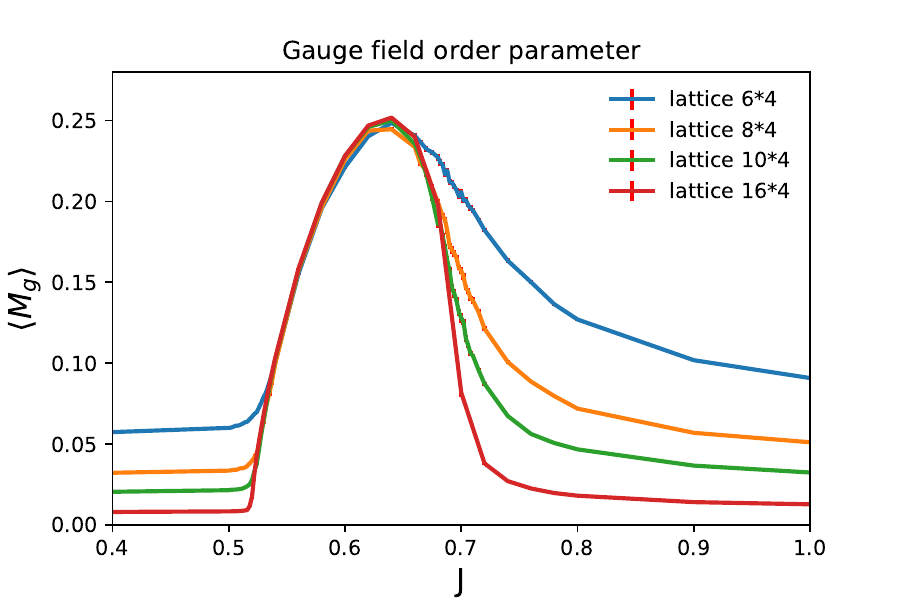} 
   \hspace*{-5mm}
   \includegraphics[width=80mm]{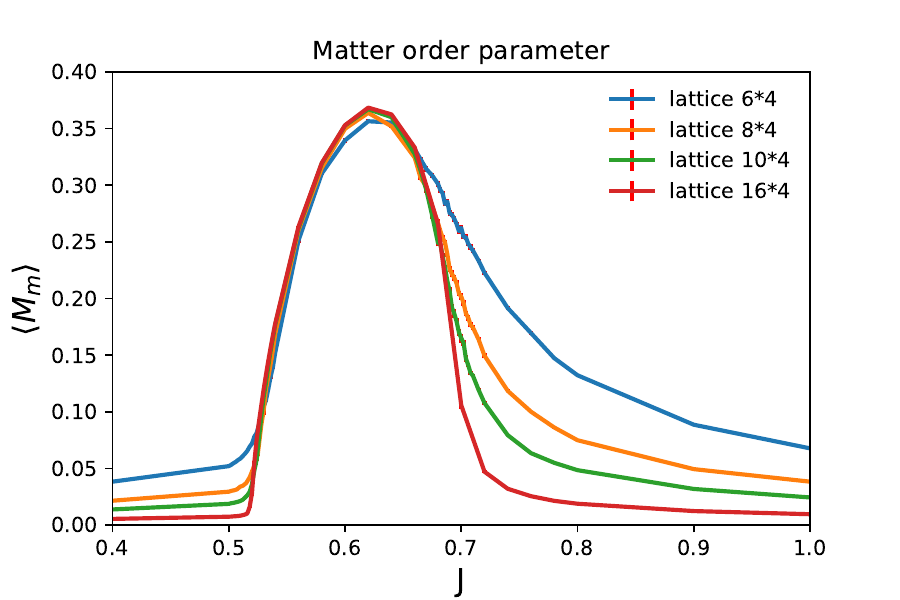} 
   \hspace*{0mm}
   \includegraphics[width=80mm]{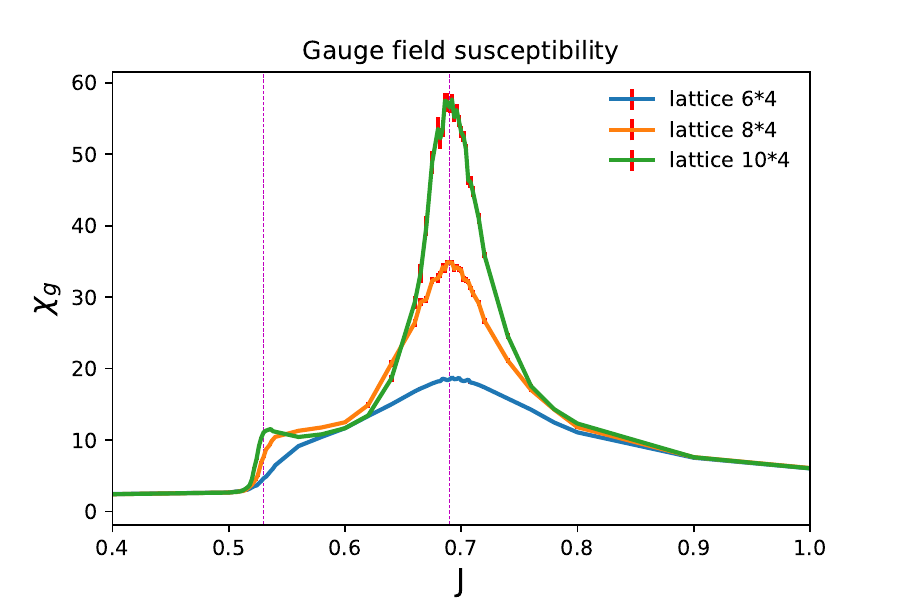} 
   \hspace*{-5mm}
   \includegraphics[width=80mm]{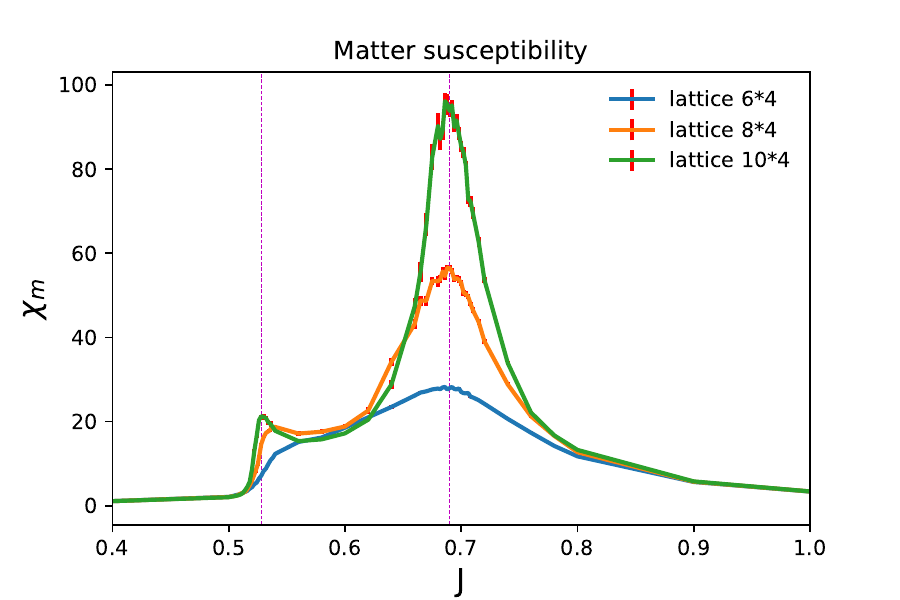} 
    \hspace*{2.5mm}
    \includegraphics[width=80mm]{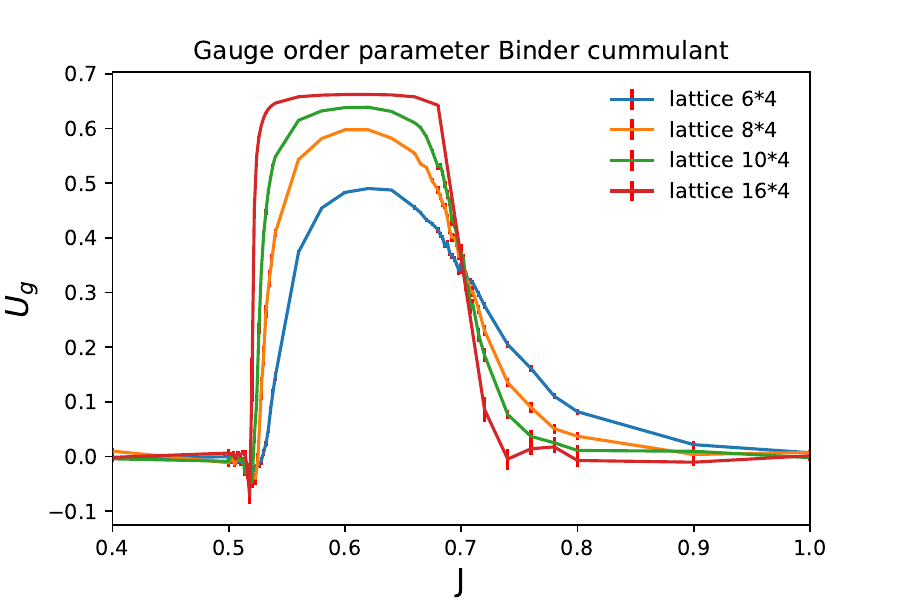} 
   \hspace*{-0.5mm}
   \includegraphics[width=80mm]{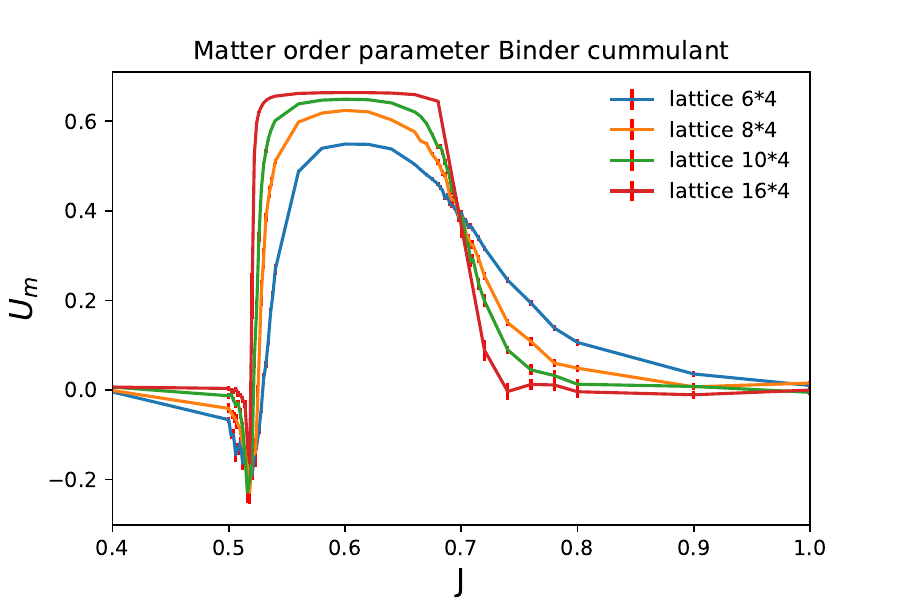} 
   \caption{Vacuum expectation value (top row), susceptibility (middle row) and Binder cumulant (bottom row) 
   for the gauge field order parameter $M_g$ (lhs.\ column) and the matter order parameter $M_m$ (rhs. column). The results are for 
   the self-dual value $\beta = \beta^* \equiv 1/2\pi$ and are plotted as a function of $J$. We compare different volumes as indicated in the legends.}
   \label{fig:overview}
\end{figure}

In Fig.~\ref{fig:overview} we show our results for the order parameters $\langle M_g \rangle$, $\langle M_m \rangle$ (top row),
for the susceptibilities $\chi_g$, $\chi_m$ (middle row) and for the Binder cumulants $U_g$ and $U_m$ (bottom row). 
The lhs.\ column shows the result for the respective gauge field quantities, while the rhs.\ column is for the matter fields. 
The observables are plotted as a function of $J$ and were determined for volumes $4^4$, $6^4$, $8^4$ and $10^4$ 
at fixed gauge coupling $\beta = \beta^* = 1/2\pi$, i.e., the self-dual value. 

All observables suggest that there is indeed spontaneous symmetry breaking as a function of $J$ with endpoints located at 
$J_1 \sim 0.52$ and $J_2 \sim 0.7$. Below $J_1$ and above $J_2$ the order parameters $\langle M_g \rangle$ and 
$\langle M_m \rangle$ approach 0 in the infinite volume limit, while inside the interval $(J_1, J_2)$ they remain finite. 
The corresponding susceptibilities develop peaks near $J_1$ and $J_2$ that scale with the volume. Finally the Binder 
cumulants allow for a
first assessment of the nature of the endpoints: Near $J_1$ they develop minima which hints at a first order endpoint 
at $J_1$. At $J_2$ the Binder cumulants for the different volumes intersect in a common point which indicates 
second order behavior at the endpoint $J = J_2$. We will study the nature of the two endpoints in more detail in the 
next subsection. 

\begin{figure}[t] 
   \hspace*{0mm}
   \includegraphics[width=80mm]{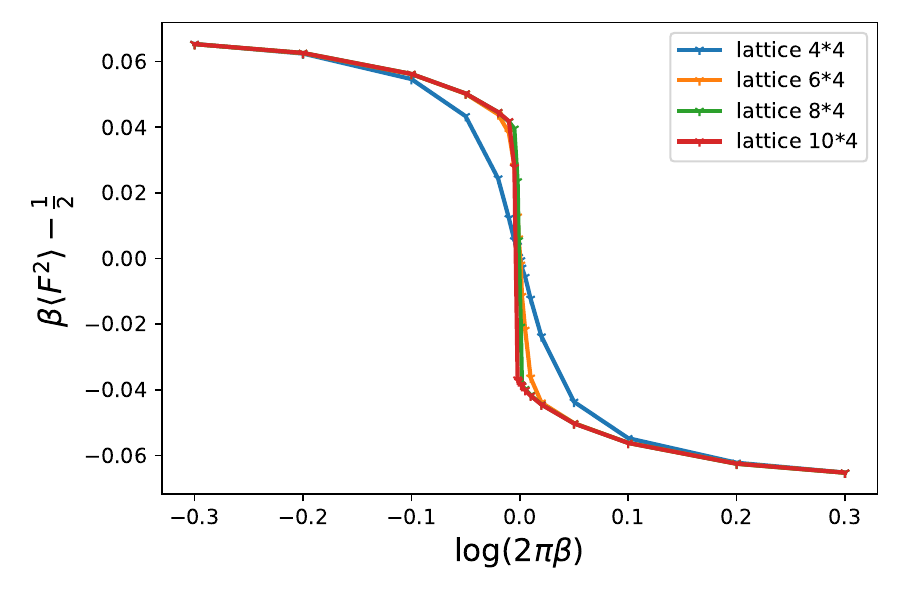} 
   \hspace*{2mm}
   \includegraphics[width=80mm]{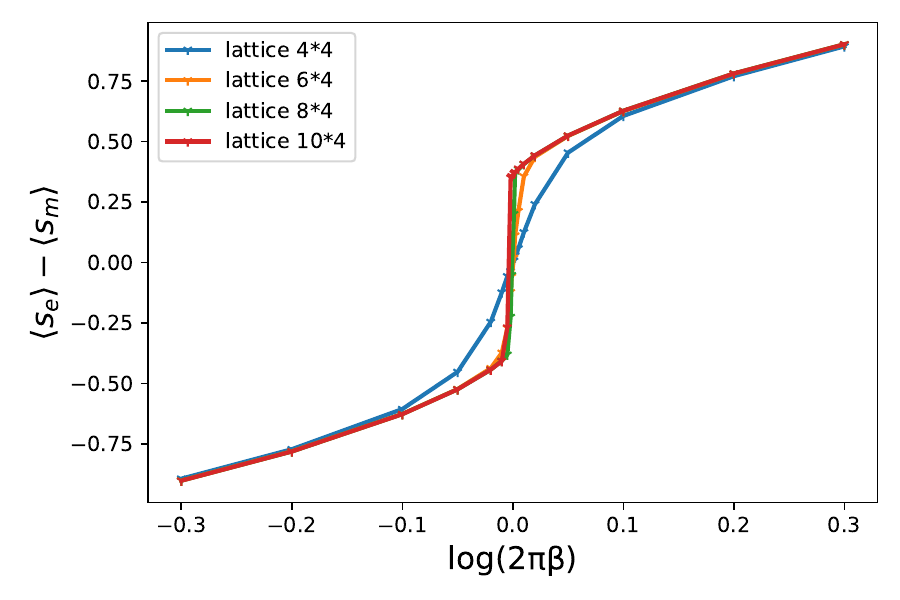} 
   \caption{The gauge field order parameter (lhs.) and the matter order parameter (rhs.) as a function 
   of the rescaled gauge coupling $\ln (2\pi \beta)$, for values across the self-dual value $\beta^* \equiv 1/2\pi$.
   The matter field coupling is set to $J = 0.6$ here.}
   \label{fig:orderparam_J06}
\end{figure}

As already announced, we also want to cross-check the nature of the transition when vertically crossing the critical 
line in the phase diagram Fig.~\ref{fig:phase_diagram}. For this study we now keep the matter field coupling fixed at
$J = 0.6$ and vary the gauge field coupling $\beta$ in the vicinity of $\beta = \beta^*$. The corresponding plots for the 
gauge field order parameter (lhs.\ plot) and the matter order parameter (rhs.) are shown in Fig.~\ref{fig:orderparam_J06}.
In order to fully display the symmetry of the first order transition we plot the gauge field coupling on the horizontal axis in
the rescaled form $\ln (2\pi \beta)$, which is odd under duality transformations and gives 0 at the self-dual point. 
The gauge field and the matter order parameters on the vertical axes are plotted in the form
\begin{equation}
\beta \langle F^2 \rangle \; - \; \frac{1}{2} \qquad \mbox{and} \qquad \langle s_e \rangle \; - \;  \langle s_m \rangle \; ,
\end{equation}
which is a form that is again odd under duality transformations, as can be seen from 
\eqref{sumruleF2_qed} and \eqref{duality_phi2}. Since both, the rescaled coupling and the order parameters are odd under 
duality transformations, the plots for the order parameters must be antisymmetric functions of $\ln (2\pi \beta)$. This is indeed what we observe 
in Fig.~\ref{fig:orderparam_J06}.  When comparing the different volumes we find that the observables quickly develop 
the discontinuity at $\ln (2\pi \beta) = 0$, which is the expected first order signature when driving the symmetry breaking coupling 
$\beta$  through the self-dual point. Thus we confirm that the vertical line in the phase diagram  Fig.~\ref{fig:phase_diagram} is indeed of
first order.

\subsection{Analysis of the endpoints}\label{sec:endpoints}

The first round of analysis in the previous subsection suggested that the endpoint at 
$J_1 \sim 0.52$ is first order, while the one at $J_2 \sim 0.7$ is of second order. In this subsection we now 
aim at determining more precisely the values of $J_1$ and $J_2$ and at characterizing the two endpoints.

\begin{figure}[t] 
   \hspace*{0mm}
   \includegraphics[width=80mm]{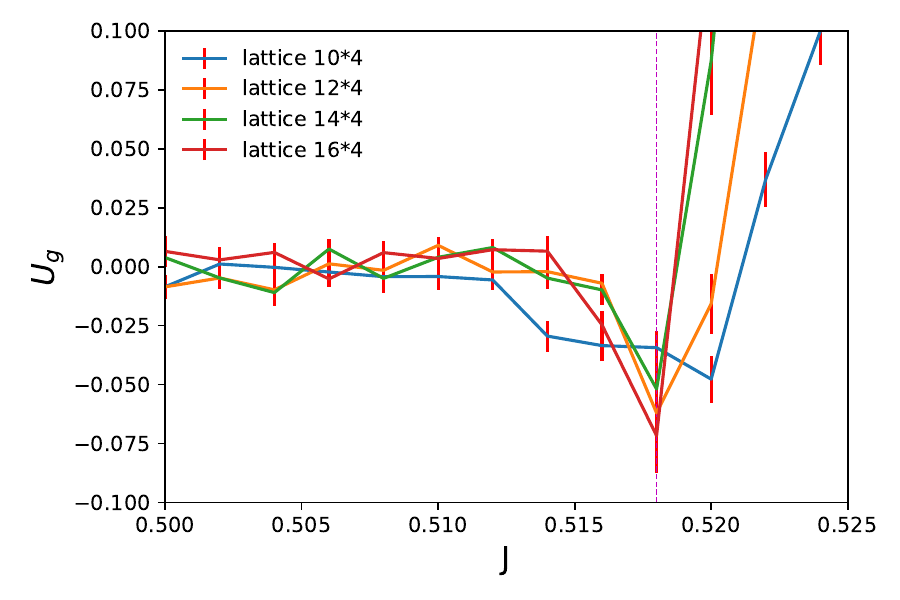} 
   \hspace*{2mm}
   \includegraphics[width=80mm]{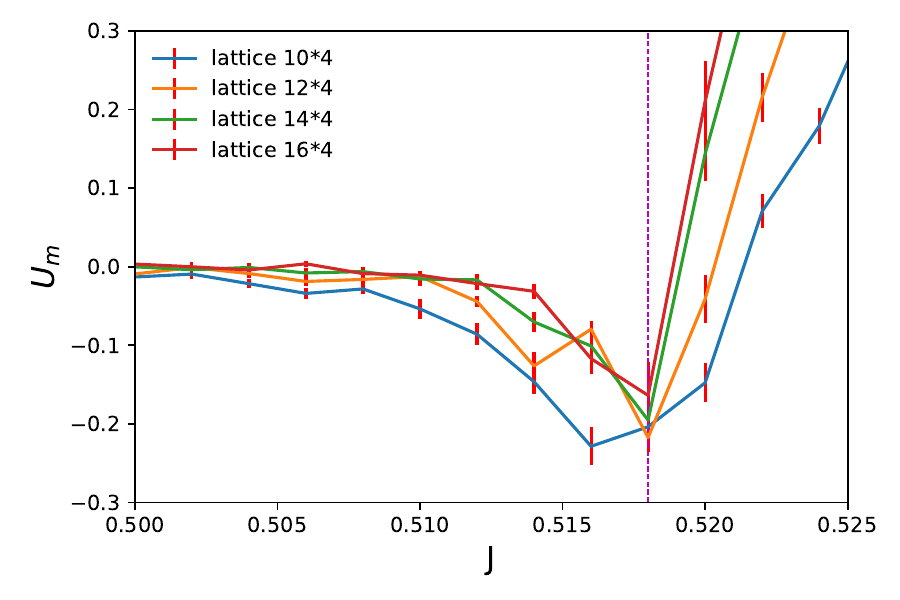} 
   \caption{The Binder cumulant 
   for the gauge field order parameter (lhs.) and the matter order parameter (rhs.). For different volumes we 
   show the results at $\beta = \beta^*$ as a function of $J$ and zoom into the region near the
   first order transition point at $J_1 \sim 0.52$.}
   \label{fig:Binder_1}
\end{figure}

For the transition at $J_1$ we observed the formation of minima in the two Binder cumulants. The positions of the minima 
converge towards the true value $J_1$ when increasing the volume. Since in the two bottom plots of Fig.~\ref{fig:overview} 
this is a little hard to see, in Fig.~\ref{fig:Binder_1} we zoom into the region near $J_1$. We find that both Binder cumulants 
form minima and that for all volumes except for the smallest volume $10^4$ the positions of the minima agree. Thus we conclude 
that we find a first order transition at $J_1 = 0.518(2)$, where the error is given by the stepsize in $J$ we use,
i.e., $\Delta J = 0.002$.

\begin{figure}[t] 
   \hspace*{5mm}
   \includegraphics[width=80mm]{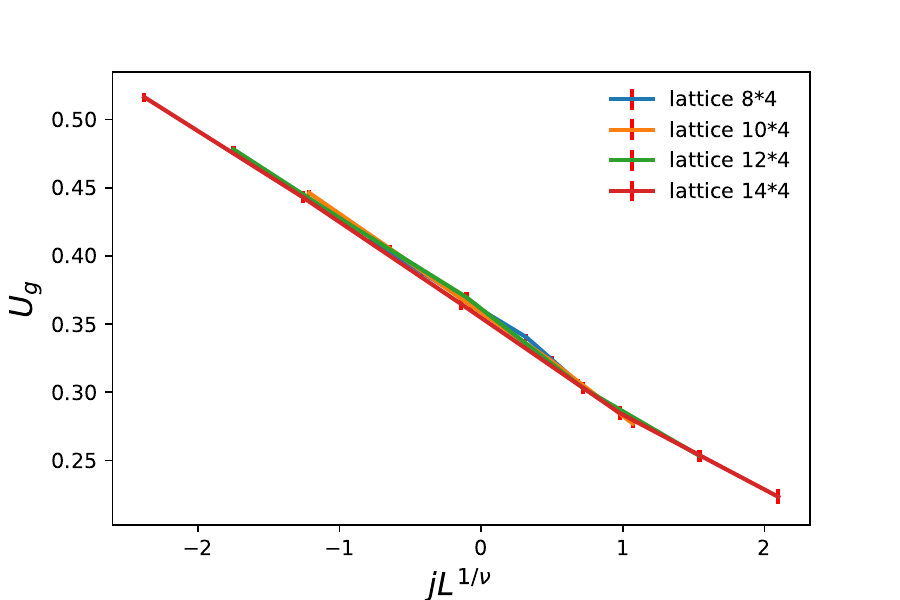} 
   \hspace*{1mm}
   \includegraphics[width=80mm]{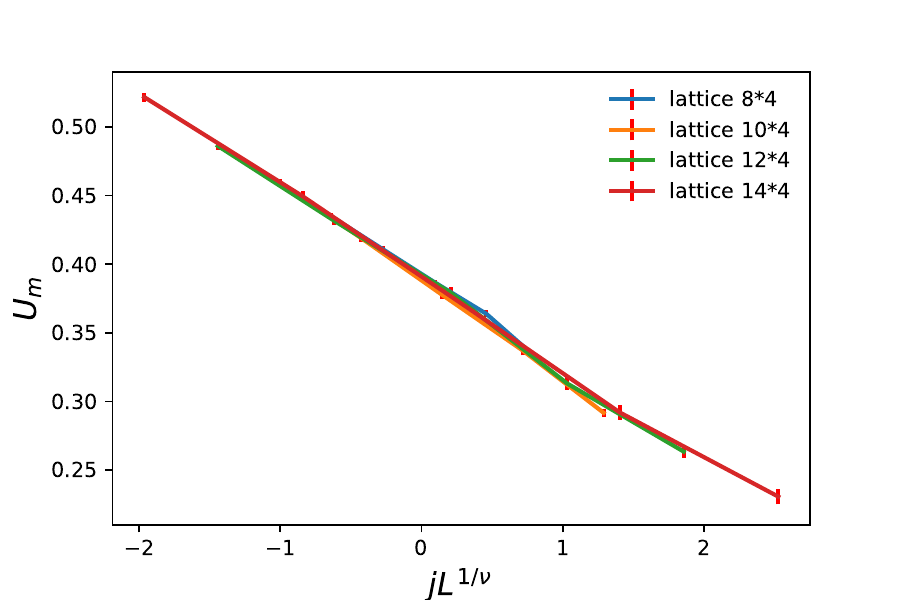} 
   \caption{The Binder cumulant 
   for the gauge field order parameter (lhs.) and the matter order parameter (rhs.). We plot  
   the results at $\beta = \beta^*$ as a function of $j \, L^{1/\nu}$ with $j = J - J_2$ 
   and $\nu = 1/2$. We compare the results for different volumes $L^4$.}
   \label{fig:Binder_2}
\end{figure}

For the transition at $J_2 \sim 0.7$  Fig.~\ref{fig:overview} suggests that the transition is of second order. In
that case the Binder cumulants are expected to obey the finite size scaling formula 
\begin{equation}
U \; \sim \; A \; + \; B \, L^{1/\nu} \, ( J - J_2) \; ,
\label{Binder_scaling}
\end{equation}
where $A$ and $B$ are constants and $\nu$ is the critical exponent for the scaling of the correlation length. Thus, when
plotting the Binder cumulants as a function of $L^{1/\nu} \, ( J - J_2)$ the results for different volumes $V = L^4$ should 
collapse to universal straight lines -- given that $\nu$ and $J_2$ are chosen correctly. As discussed in Subsection \ref{sec:phase_diagram},
we conjecture that the transition is in the 4d Ising universality class, i.e., a Gaussian fixed point, such that we expect $\nu = 1/2$. We test 
this hypothesis by setting $\nu = 1/2$ in the scaling formula \eqref{Binder_scaling} and in Fig.~\ref{fig:Binder_2} plot the Binder 
cumulants as function of $L^2 \, (J - J_2)$, where $J_2$ is treated as a free parameter which we choose such that we find the best 
collapse of the data for different $L$. The lhs.\ plot in Fig.~\ref{fig:Binder_2} shows the results for the gauge field Binder cumulant,
while the rhs.\ is for the matter Binder cumulant. In both cases the collapse confirms the expected critical exponent $\nu = 1/2$ and
the critical coupling is determined to be $J_2 = 0.700(1)$. 

\begin{figure}[t] 
   \hspace*{5mm}
   \includegraphics[width=80mm]{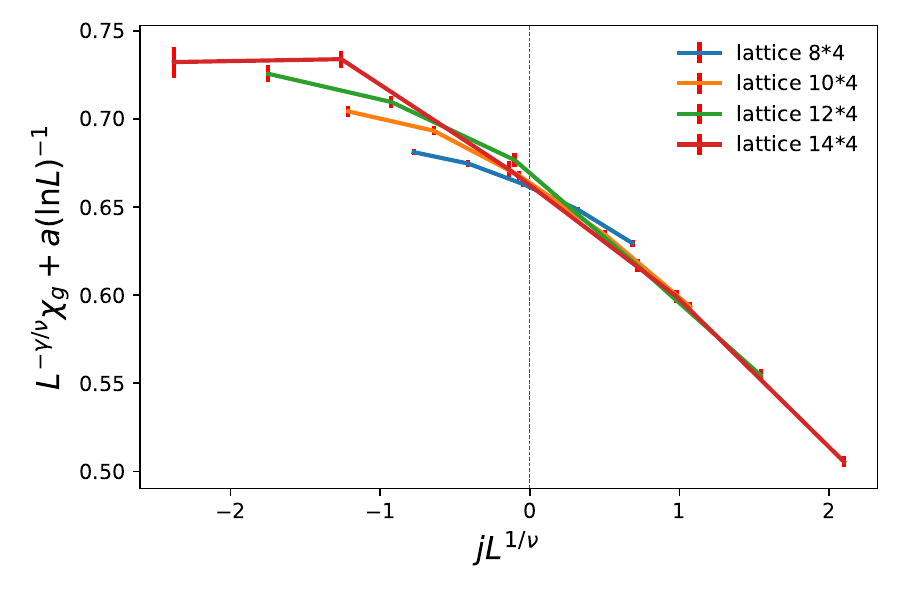} 
   \hspace*{1mm}
   \includegraphics[width=80mm]{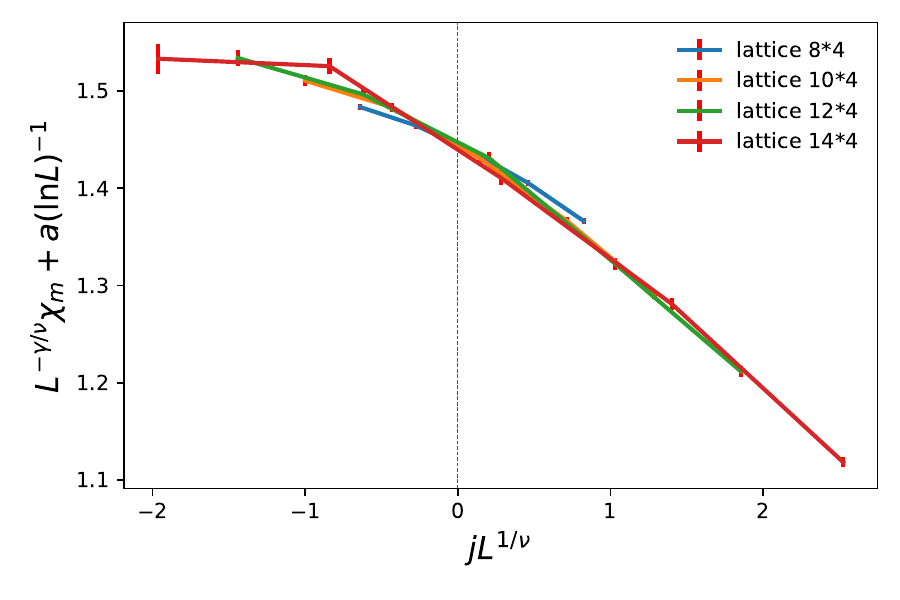} 
 \caption{Finite size scaling analysis of the gauge field (lhs.\ plot) and matter (rhs.) susceptibilities.}
   \label{fig:chi_fss}
\end{figure}

Finally we also try to confirm the universality class of the second order transition by analyzing the susceptibilities and the corresponding critical
exponent $\gamma$, which for the conjectured Gaussian fixed point would be $\gamma = 1$. Again we employ finite size scaling which for the 
susceptibilities takes the form  (we define $j \equiv J - J_2$),
\begin{equation}
\chi \; \sim \; L^{\gamma/\nu} \left( g(j L^{1/\nu}) + \frac{f(j L^{1/\nu})}{\ln(L)} \right) \; ,
\end{equation}
where here also the leading logarithmic corrections are taken into account. Considering only the constant term in 
the expansion of the scaling function $f$, i.e., setting $f = - a$, where $a$ is some constant we find that $\chi \, L^{- \gamma/\nu} + a / \ln(L)$ 
should be a universal function of $j L^{1/\nu}$ for a suitably chosen parameter $a$. In Fig.~\ref{fig:chi_fss} we plot the combination 
$\chi \, L^{- \gamma/\nu} + a / \ln(L)$ with $\gamma = 1$ as a function of $j L^{1/\nu}$ for different volumes $L^4$. The lhs.\ plot  
is for the gauge field susceptibility, while the rhs. shows the results for the matter susceptibility. The parameter $a$ was chosen such that 
the collapse of the data for the different volumes is optimized. We find that the collapse is not as good as for the Binder cumulants, but 
nevertheless confirm, that a value of $\gamma = 1$ is plausible. 
We may summarize the discussion of the endpoints as follows: At $J_1 = 0.518(2)$ we find a first order transition, while at $J_2 = 0.700(1)$ the
transition is of second order with critical exponents that are compatible with the conjectured 4d Ising universality class, i.e., a Gaussian fixed point.

We finally explore the possibility expressed in Subsection 2.3 that charge conjugation $\C$ might be broken in the vicinity of the horizontal line of the 
phase diagram sketched in Fig.~\ref{fig:phase_diagram}. For this study we consider the following two order parameters that are odd under $\C$, 
which in continuum language are defined as
\begin{equation}
T_1 \; \sim \; \sum_\mu  j^{e}_\mu \, \partial_\mu \sum_{\alpha < \beta} F_{\alpha \beta} F_{\alpha \beta} 
\qquad \mbox{and} \qquad 
T_2 \; \sim \; \sum_\mu  j^{e}_\mu \, \partial_\mu \sum_{\alpha < \beta \atop \rho < \sigma} \, 
\epsilon_{\alpha \beta \rho \sigma} \, F_{\alpha \beta}  \, F_{\rho \sigma} \; .
\end{equation}
Here $j^{e}_\mu$ is the current of the electric matter which we discretize as
$j^{e}_{x,\mu} = \, \mbox{Im} \, \phi^{e\, *}_x e^{\, i A_{x,\mu}^e} \, \phi_{x+\hat \mu}^e$. For the discretization of the field strength
we use \eqref{eq:F_def} and the partial derivative $\partial_\mu$ is discretized with a nearest neighbor difference.

\begin{figure}[t] 
   \hspace*{6.5mm}
   \includegraphics[width=85mm]{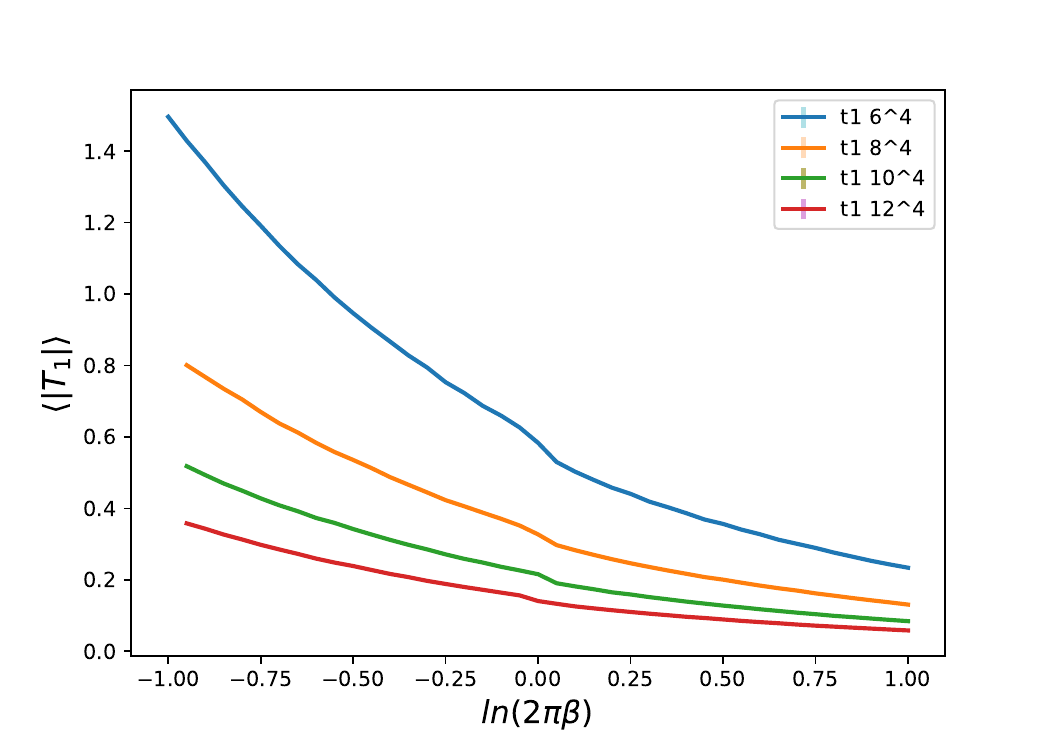} 
   \hspace*{-5mm}
   \includegraphics[width=82mm]{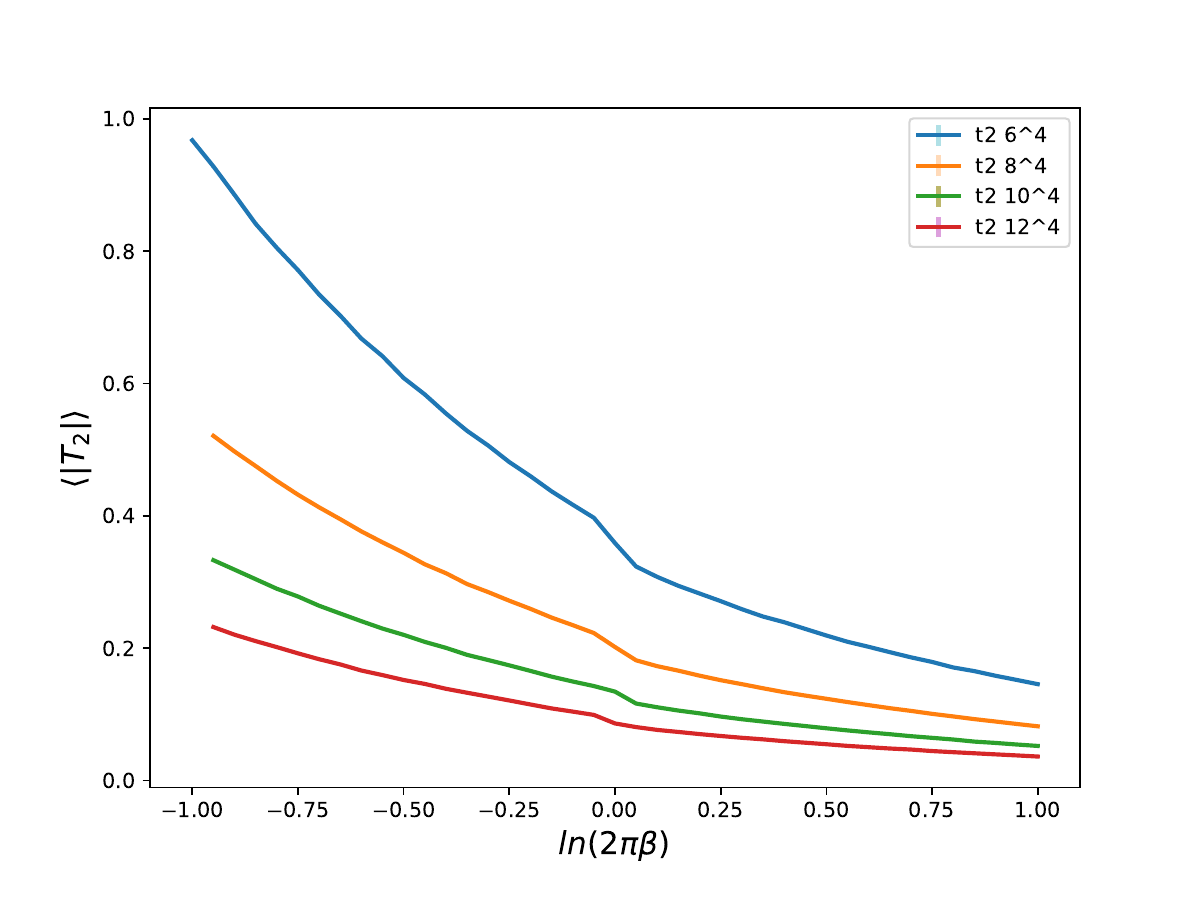} 
    \hspace*{6.5mm}
   \includegraphics[width=85mm]{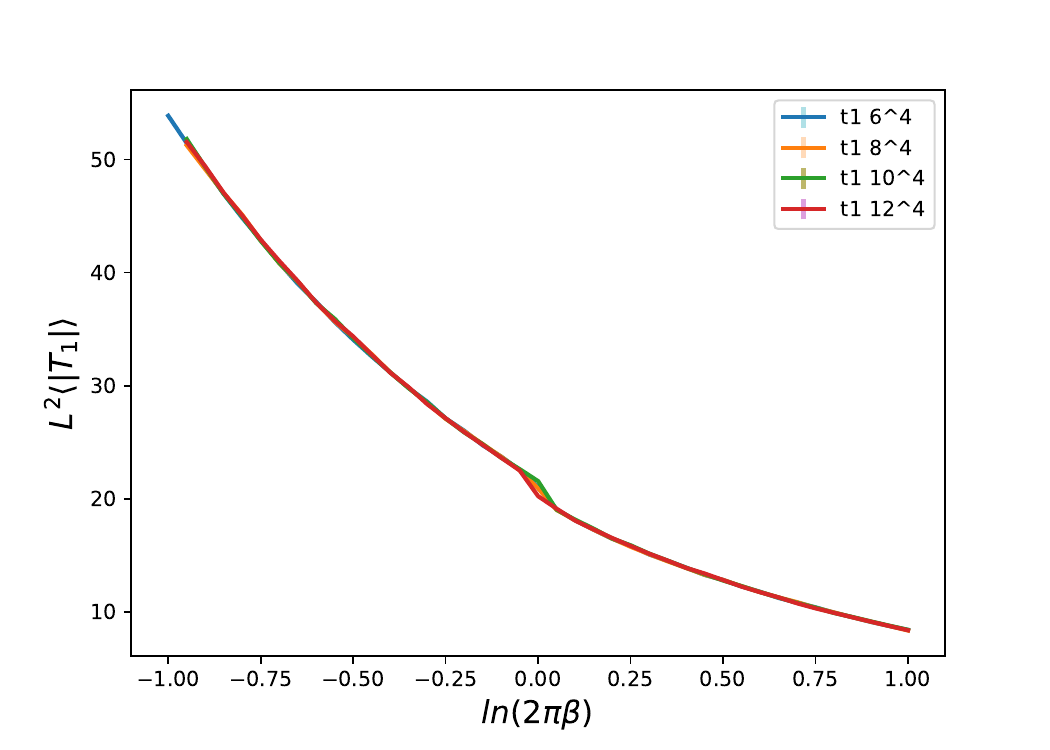} 
   \hspace*{-5mm}
   \includegraphics[width=82mm]{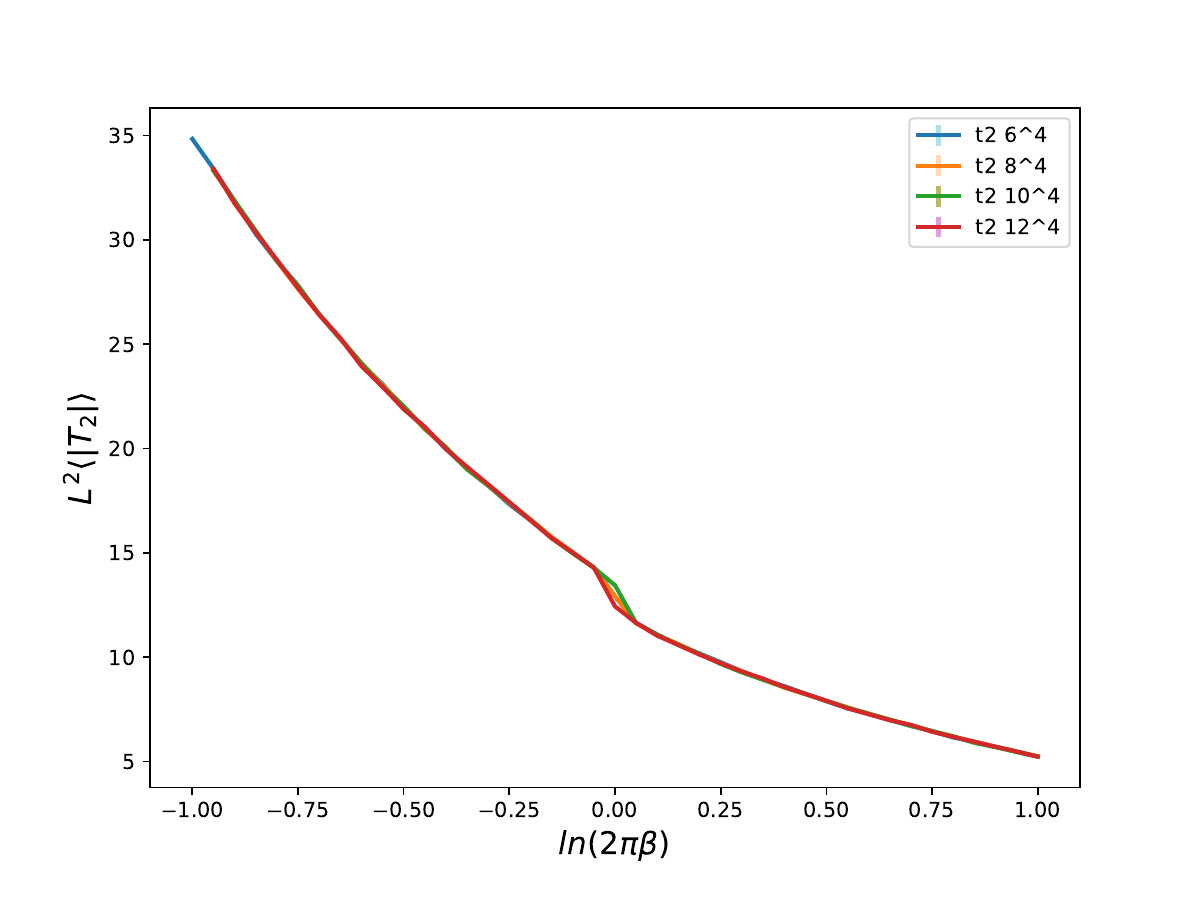} 
   \caption{Results for the $C$-breaking order parameters $T_1$ and $T_2$. In the top row of plots we show
   $\langle | T_1 | \rangle$ (lhs.) and $\langle | T_2 | \rangle$ (rhs.) for different volumes $L^4$ with $L = 8,10,12,14$.
   The results are plotted as a function of $\ln (2\pi \beta)$ and we use a matter field coupling of $J = 0.65$.
   In the bottom row of plots we show the rescaled expectation values $L^2 \, \langle | T_1 | \rangle$ (lhs.) and 
   $L^2 \, \langle | T_2 | \rangle$ (rhs.), illustrating that the expectation values vanish as $1/L^2 = 1/\sqrt{V}$. 
   All error bars are smaller than the line width.}
   \label{fig:t1t2}
\end{figure}

In Fig.~\ref{fig:t1t2} we show our results for $\langle | T_1 | \rangle$ and $\langle | T_2 | \rangle$ in the top row of plots, 
while at the bottom the two expectation values are rescaled with $L^2$. We work at a fixed matter coupling of $J = 0.65$ 
and compare four different volumes $L^4$ with $L = 8,10,12$ and $14$. The results are again plotted as a function of $\ln(2 \pi \beta)$.
Note that the order parameters are not symmetrized under ${\cal S}$, such that here we do not expect symmetry under 
$\beta \leftrightarrow \tilde \beta$.

The top row of plots show that both expectation values  $\langle | T_1 | \rangle$ and $\langle | T_2 | \rangle$ vanish in the
thermodynamic limit. In the bottom plots we show $L^2 \, \langle | T_1 | \rangle$ (lhs.) and 
 $L^2 \, \langle | T_2 | \rangle$ (rhs.). This rescaling collapses the data and establishes that the volume scaling is 
 $\propto 1/L^2 = 1/\sqrt{V}$, including also the small remnant of the first order transition at $\ln(2\pi \beta) = 0$,
 i.e., at the self dual point $\beta = 1/2\pi$. 
 
 The $1/\sqrt{V}$  behavior
 may be understood be viewing the values for $T_1$ and $T_2$ at some space-time point as independent random variables. 
 The distribution of the average of random variables with sample size $N$ has a standard devation scaling as $1/\sqrt{N}$.
 In our case the values of the order parameters $T_1$ and $T_2$ on the $V$ space-time points are not fully independent, 
 but since there is a mass-gap, $N$ is roughly given by the volume $V = L^4$. This gives rise to the $1/\sqrt{V}$ scaling we observe. 
 We conclude that the order parameters $T_1$ and $T_2$ have vanishing expectation value in the thermodynamic limit and 
 $\C$ remains unbroken.

\section{Conclusion and future prospects}\label{sec:conclusion}

In this work we discussed the possible phase structure of self-dual U(1) lattice gauge theories based on a modified Villain action. 
We have seen that the space of such theories is large, allowing arbitrary matter to be coupled electrically and/or magnetically. 
When coupling multiplets of such matter fields the electric and magnetic flavor symmetry that arises often has 't Hooft anomalies, 
eliminating the possibility of a trivially gapped phase. 

An interesting question is whether any of such theories have a self-dual CFT fixed point. The natural place to look for the new fixed 
points is the self-dual line $\beta=1/2\pi$ as in that case the coupling is protected by self-duality. For a single electric and magnetic bosonic flavor $N_f^e=N_f^m=N_f=1$ we argued, 
and numerically confirmed, that the phase structure along the self-dual line has two transitions as the bosonic matter condenses. 
The transition from the photon phase to the Higgs/confined phase is 1st order, continuing in the Higgs/monopole coexistence phase, 
which breaks the self-dual symmetry spontaneously. The coexistence phase then disappears in a 2nd order 4d Ising (i.e. gaussian) transition, and a 
trivially gapped phase ensues. 

For the case of $N_f>1$ the bosonic theory has a mixed 't Hooft anomaly between the two flavor symmetry groups 
$PSU(N_f)\times PSU(N_f)$ and a trivial phase is not allowed. Still, the transition between the photon phase and the self-duality 
broken phase is likely 1st order, because away from the self-dual point for a weak electric (magnetic) coupling this can be shown 
by perturbative RG flow equations. However, for $N_f>183$ the RG equations allow for a 2nd order transition, which may persist 
all the way to the self-dual line, resulting in an interacting fixed point. If this is the case the fixed point has to be interacting, because the electric 
coupling is fixed exactly by self-duality. 

We here also add a comment on fermionic self-dual theories. To begin with, let us consider QED without monopoles
and with $N_f$ massive electric flavors of mass $m_e$. The flavor group is $PSU(N_f)$ and if the electric coupling is 
sufficiently weak, the theory flows to a free photon phase. When the mass of the electric flavors is exactly zero, the theory 
flows to a theory of free photons and $N_f$ fermions.

Moreover, the above conclusion must follow even when the coupling is strong, as long as $N_f$ is sufficiently large. 
Namely if the coupling is strong, the RG iteration will generate a term $\propto N_f F_{\mu\nu}^2$ and render the electric 
coupling weak. It is then natural to conjecture that for any flavor $N_f$ the massless theory flows to a phase non-interacting photons 
and $N_f$ fermions. Note also that the massless limit has an $\SU(N_f)$ axial symmetry, which has a 't Hooft anomaly. The anomaly 
must be saturated, and it is saturated by the free fermions. One might wonder whether $\SU(N_f)$ could spontaneously break
and saturate the anomaly in this way. This option is certainly not eliminated, but intuitively we expect to need confinement for 
spontaneous symmetry breaking, and to generate confinement we need magnetically charged dynamically matter, which is absent.

\begin{figure}[!t] 
   \centering
   \includegraphics[width=3.5in]{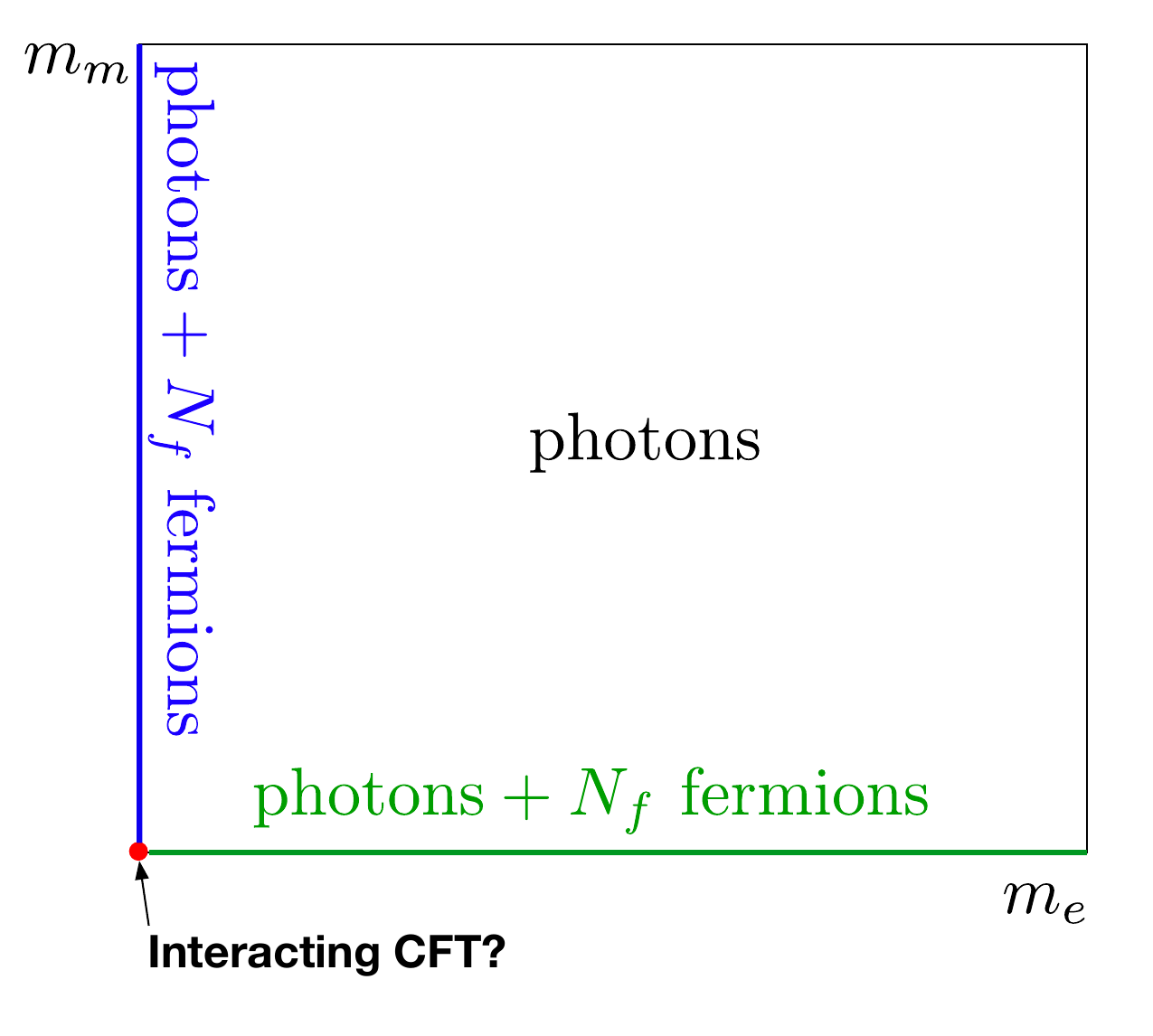} 
   \caption{The conjectured phase diagram of the theory at the self-dual coupling, with $N_f$ electric and $N_f$ magnetic fermions. 
   The point when fermions are exactly massless is likely to be an interacting CFT.}
   \label{fig:fermion_phase_diag}
\end{figure}

Now consider coupling $N_f$ fermions to the magnetic gauge field, and endow the electric and magnetic fermions with masses 
$m_e$ and $m_m$ respectively. Then, a natural conjecture is that when $m_e$ and $m_m$ are non-zero, the theory flows to free 
photons. If one of the masses is zero, then by arguments above the theory flows to free massless fermions and photons (see Fig~\ref{fig:fermion_phase_diag}). The questions 
is what happens when both electric and magnetic fermions are massless. In this case, there exists a vector symmetry $\PSU(N_f)\times\PSU(N_f)$ 
and an axial vector symmetry $\SU(N_f) \times \SU(N_f)$. As we discussed in the main text, the vector symmetries have  
mixed 't Hooft anomalies. However, the axial vector symmetries have 't Hooft anomalies individually, which must both be satisfied.

The 't Hooft anomalies are typically saturated by either spontaneously breaking the symmetries, or by a CFT. The spontaneous breaking 
of the symmetry seems unlikely to us. The reason is that a spontaneously broken phase is typically robust against small 
perturbations\footnote{Save for maybe lifting goldstones if the perturbation breaks the broken symmetry.}. Yet the massless 
point is surrounded by free CFTs, which suggests that the massless point is a CFT itself. Unfortunately this lattice theory 
cannot be simulated easily, because of the fermionic sign-problem.

However, for both bosonic and fermionic self-dual theories, one can try to bootstrap the tentative CFT, imposing the symmetries 
and 't Hooft anomalies that we found. The bootstrap approach may yield insights into whether such CFT phases are expected 
at small numbers of flavors, as well as into the properties of these CFTs. Another interesting approach is to attempt to build 
fermionic models which evade the sign problem and simulate them.

\vskip20mm
\noindent
{\bf Acknowledgments:} 
We thank Madalena Lemos, John McGreevy, Mithat \"Unsal and Yuya Tanizaki for discussions. TS is supported by the Royal Society University Research Grant. NI is supported in part by by the STFC under consolidated grant ST/L000407/1
 
\newpage
\begin{appendix}

\section{The RG equations}\label{app:RG_solution}

\subsection{Setup}

In this section, we discuss the RG flow of $N_{f}$ complex scalars coupled to a $\U(1)$ abelian gauge field, i.e., the theory described by the action \eqref{actionNfscalars}, which we reproduce here for convenience:
\be
S[\phi,A] = \int d^4x \le(\frac{1}{4e^2} F^2 + \sum_{i=1}^{N_f} \le(|\p \phi_i|^2 + m^2 |\phi_i|^2\ri) + \frac{\lambda}{4} \le(\sum_{i=1}^{N_f}|\phi_i|^2\ri)^2\ri)
\ee 

\begin{figure}[tbp] 
   \centering
   \includegraphics[width=4in]{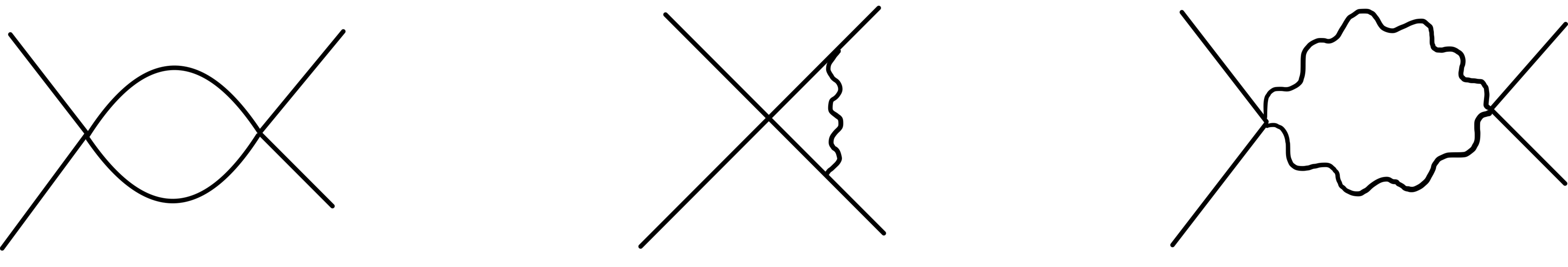} 
   \caption{The topology of the diagrams contributing to the running of the $\phi^4$-coupling.  }
   \label{fig:lambda_diagrams}
\end{figure}

We are interested in understanding the order of the phase transition at $m^2 \sim 0$ that separates the Higgsed and Coulomb phases. 
If we can generically arrive at an RG fixed point by tuning only the single parameter $m^2$, then this means that the phase transition 
is of second order. On the other hand, if we generically don't arrive at such a fixed point, then the transition will be first order. 
  
To that end, we tune the mass of the scalars to zero, and look at the RG flow equations for $g=e^2$ and $\lambda$. On general grounds they
have the form
\begin{subequations}\label{eqs:RG_flow0}
\begin{align}
&\frac{dg}{d\ell} = -a g^2\;,\\
\label{eq:RG_flow0_lam}
&\frac{d\lambda}{d\ell}= -b\lambda^2-c\lambda g-d g^2\;,
\end{align}
\end{subequations}
where $\ell$ parametrizes the RG flow, i.e., if $\Lambda$ is the mass-dimension UV cutoff of the theory, and we flow to $\Lambda'<\Lambda$, then  $\ell= \log(\Lambda/\Lambda')$.  

Let us discuss these equations a bit. The first equation is just the running of the gauge coupling which has no contributions 
from $\phi^4$ interactions to this order. Famously $a>0$, so that the coupling $g=e^2$ becomes more negative in the IR, i.e., QED is IR free. 

The diagrams contributing to the beta-function for $\lambda$ are given in the top of Fig.~\ref{fig:lambda_diagrams}. 
The first of these diagrams is of order $\lambda^2$, and should have $b>0$ (i.e., $\lambda$ is marginally irrelevant). 
The last diagram is similar, except that a photon runs in the loop. Finally the middle diagram contributes to the coefficient $c$ in (\ref{eq:RG_flow0_lam}) 
and turns out to be positive, such that $c<0$, although the right-hand side is still a negative-definite quadratic form in the $(\lambda, g)$ space. 

Let us now discuss the space of solutions to these equations. The only fixed point is at $(\lam, g) = (0,0)$. Do we 
approach this fixed point from generic initial data $(\lam, g) > 0$? As the right-hand sides of both equations are negative, 
both positive couplings $(\lam, g)$ become smaller in the IR. If there was only one coupling (e.g., $\lambda>0$), then this would imply 
that $\lambda$ flows to zero and the theory becomes IR free. However as there are two couplings, it is quite possible for the RG flow to
\emph{miss} the origin and flow in the infrared towards increasingly smaller values of $\lambda$, eventually becoming negative -- presumably corresponding to a phase 
 where the scalars are condensed\footnote{Negative $\lambda$ seems like it would destabilize the theory. However recall that all couplings of type $|\phi|^{2n}$ are generated, and are expected to stabilize the potential}. Whether or not this generically happens requires a more careful study of the RG equations. 

\begin{figure}[htbp] 
   \centering
   \includegraphics[width=3.5in]{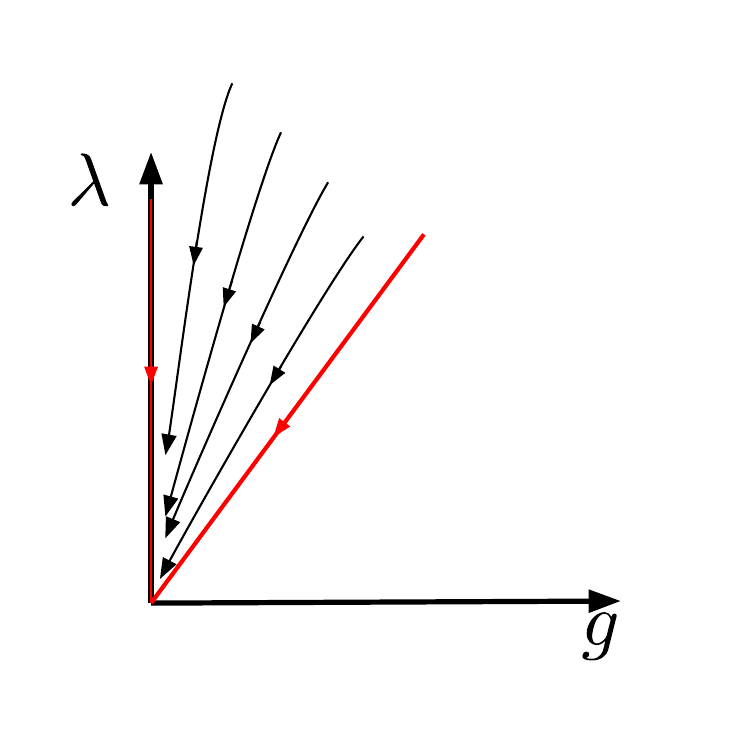} 
   \caption{The RG flow is bounded by the separatrix solutions $g=\alpha\lambda$ with $\alpha$ given in \eqref{eq:separatrices} and drawn as red lines. The separatrices form a funnel which guaranties that the fixed point is reached as long as the flow is started within the funnel.}
   \label{fig:funnel}
\end{figure}

To do this, we will seek to construct {\it separatrices} in solution space, i.e., one-dimensional lines $L \subset (\lam, g)$ 
running through the origin such that if the system is started with initial conditions on $L$, under RG flow the system remains 
on $L$ for all RG time. To find $L$, we just simultaneously solve the following equations:
\be
g = \alpha \lambda \qquad \frac{d g}{d \ell} = \alpha \frac{d \lambda}{d\ell}
\ee
with $\alpha$ a constant that is the slope of the line. These have the solutions
\be\label{eq:separatrices}
\alpha = 0 \qquad \mbox{and} \qquad \alpha = \frac{1}{2d}\le(a-c \pm K\ri) \quad , \quad K\equiv\sqrt{(a-c)^2-4db} \; .
\ee
Thus, if $K$ is positive, we may construct real separatrices along which the RG flow definitely hits the origin. 
Because solutions to a first order differential equation cannot cross, this means that all points inside a funnel between 
two seperatrices will also  hit the origin. We conclude that if $K > 0$ there exists an open set of initial data from which 
we reach the free fixed point by tuning only a single parameter $m^2$, and thus that the phase transition is second order. 
Otherwise, the RG flows cannot be bounded, the couplings flow to negative infinity, and we expect the transition to be first order.
This is borne out by an exact solution below (see Fig.~\ref{fig:funnel}).

To proceed, we need values for $(a,b,c,d)$, which are given by \eqref{eqs:RG_flow1} repeated here for convenience

\begin{align}
&\frac{dg}{d\ell}=-\frac{N_fg^2}{3}\;,\tag{\ref{eq:RG_flow1g}}\\
&\frac{d\lambda}{d\ell}=-(N_f+4)\lambda^2+6\lambda g-6 g^2\;.\tag{\ref{eq:RG_flow1_lam}}
\end{align}

Let us briefly discuss the terms entering the equations above. The beta function appearing on the r.h.s. of the equation for $g$ is the vacuum polarization, which is enhanced by $N_f$. The beta function of $\lambda$ has three contributions given in Fig.~\ref{fig:lambda_diagrams}. The latter two diagrams do not get an $N_f$ enhancement, but the first diagram has an enhancement by $N_f+4$ because the flavors can run in a loop. 

Putting in these values we see that $K > 0$ if $N_{f} > 182.95$. Thus only if we have more than 183 flavors is the transition second order. 

Finally, we note that in this case we can also exactly solve the RG flow equations. In the remainder of this appendix we write out the exact solution.

\subsection{The RG analysis of scalar QED}\label{app:RG_analysis} 

Beginning with the general equations \eqref{eqs:RG_flow0}, we can define a new coupling $\bar\lambda=\lambda+Ag$, 
where $A$ is a constant chosen such that the $g^2$ term on the rhs.\ of \eqref{eq:RG_flow0_lam} vanishes. A simple calculation yields
\be
\bar\lambda=-b\bar\lambda^2-(c-2Ab)\bar\lambda g-\Big(bA^2+(a-c)A+d\Big)g^2\;.
\ee
The $g^2$ term vanishes if we set
\be
bA^2+(a-c)A+d \; = \; 0 \;,
\ee
such that
\be
A_\pm = \frac{-(a-c)\pm \sqrt{(a-c)^2-4db}}{2b}\;.
\ee
The RG equations then take the form
\begin{subequations}\label{eqs:RG_flow2}
\begin{align}
&\frac{d\bar g}{d\ell} = -\bar a \bar g^2\;,\\
\label{eq:RG_flow2_lam}
&\frac{d\bar\lambda}{d\ell}= -\bar b\bar\lambda^2-\bar c\bar \lambda \bar g\;,
\end{align}
\end{subequations}
where
\begin{align}
&\bar g(\ell)=g(\ell)\;, &&\bar \lambda(\ell) = \lambda(\ell)+A_\pm g(\ell)\;,\\
&\bar a= a\;, &&\bar b= b\;,\\
&\bar c= c-2Ab\;.
\end{align}
The solution of \eqref{eqs:RG_flow2} can be found explicitly (see Appendix~\ref{app:RG_solution}), which we can use to 
write down the general solution for $\lambda(\ell)$ and $g(\ell)$.
\begin{subequations}\label{eqs:RG_solution0}
\begin{align}
& g(\ell)= \frac{g_0}{1+ag_0\ell}\;,\\
& \lambda(\ell)= -\frac{( c- a-2A_\pm b) g(\ell)^{\frac{ c-2A_\pm b}{ a}}}{ b\left(g(\ell)^{\frac{c- a-2b A_\pm}{a}}-g_0^{\frac{ c- a-2A_\pm b}{ a}}\right)-( c- a-2A_\pm b)\frac{g_0^{\frac{ c-2A_\pm b}{ a}}}{\lambda_0+A_\pm g_0}}+A_\pm g(\ell)\;.
\label{eq:lambda_sol0}
\end{align}
\end{subequations}
The equation for $\lambda(\ell)$ can be rewritten as
\begin{subequations}\label{eqs:RG_solution}
\begin{align}\label{eq:g_sol}
& g(\ell)= \frac{g_0}{1+ag_0\ell}\;,\\
\label{eq:lambda_sol}
& \lambda(\ell)= -g(\ell)\frac{(\mp K) g(\ell)^{\frac{ \mp K}{ a}}}{ b\left(g(\ell)^{\frac{\mp K}{a}}-
g_0^{\frac{\mp K}{ a}}\right)\pm K\frac{g_0^{\frac{ \mp K}{ a}}}{\frac{\lambda_0}{g_0}+A_\pm}}+A_\pm g(\ell)\;,
\end{align}
\end{subequations}
where as above,
\be
K=\sqrt{(a-c)^2-4db} \; .
\ee
Note that both signs used above give the same solution.

If $(a-c)^2-4db<0$ then $K=i\kappa$, with $\kappa\in \mathbb R$, and the solution must be oscillatory in $\ell$, 
because it depends on the combination $g(\ell)^{ K/a}=e^{i\frac{\kappa}{a}\log(g(\ell)}$. On the other hand we expect 
$\lambda(\ell)$ to be monotonically decreasing. Indeed we can write
\be
\frac{d\lambda}{d\ell} \; = \; - \,b(\lambda+\frac{c}{2b})^2+\frac{c^2-4bd}{4b}g^2\;,
\ee
and as long as $c^2-4bd<0$ holds $\lambda$ must be monotone decreasing. 
This condition is satisfied for all $N_f$ in equation \eqref{eqs:RG_flow0}. 

Since the function $\lambda(\ell)$ has to be both oscillatory and monotone, it must be singular at finite time 
$\ell$ and $\lambda$ flows to minus infinity at finite $\ell$, which means that the fixed point is not reached and the transition is 1st order. 

On the other hand if $(a-c)^2-4db>0$, then $K$ is real and the solution $\lambda(\ell)$ is no longer oscillatory, hence there is 
no obstruction to the flow reaching the fixed point $\lambda=g=0$. This condition is fulfilled if 
$N_f> 182.95$\footnote{Note that this is precisely the condition for the existence of an interacting Wilson-Fischer fixed point in the 
$\epsilon$-expansion \cite{Halperin:1973jh}.}. What needs to be checked is that the denominator appearing in the solution does 
not go to zero for any value of $\ell$. 

\begin{figure}[htbp] 
   \centering
   \includegraphics[width=3.7in]{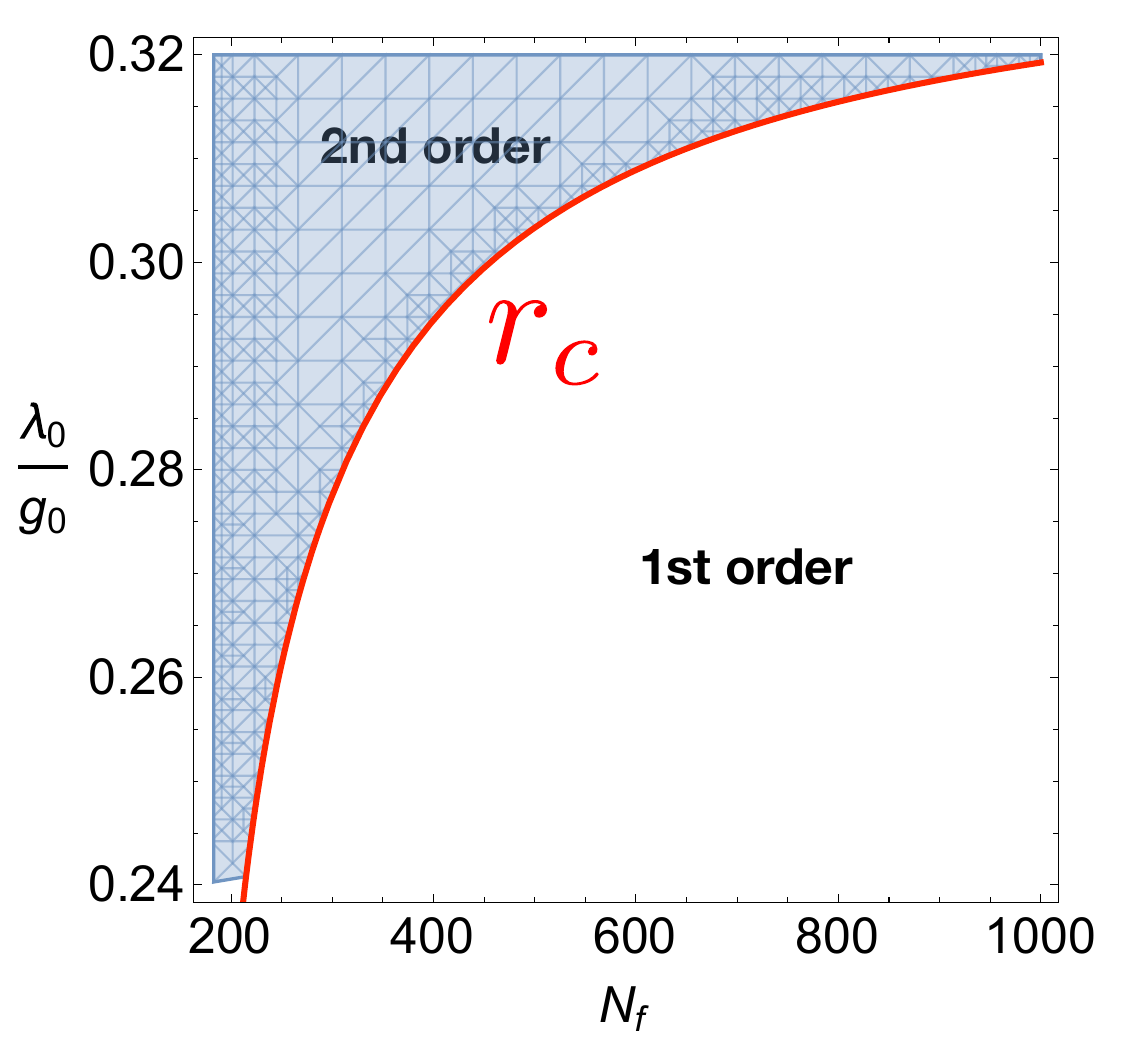}
   \caption{The plot of the critical ratio $r_0$ as a function of $N_f$ defined in \eqref{eq:critical_ratio}. For the ratio of the 
   bare couplings $\lambda_0/g_0$ in the shaded region the transition is 2nd order. }
   \label{fig:critical_ratio}
\end{figure}

We take the lower sign in \eqref{eqs:RG_solution} and examine the possibility that the denominator is zero, i.e.
\be
b\left(g(\ell)^{\frac{K}{a}}-g_0^{\frac{ K}{ a}}\right)- K\frac{g_0^{\frac{  K}{ a}}}{\frac{\lambda_0}{g_0}+A_-} \; = \; 0\;,
\ee
such that
\be
g(\ell)^{\frac{K}{a}}=g_0^{\frac{K}{a}}\left(1+\frac{K}{b}\frac{1}{\frac{\lambda_0}{g_0}+A_-}\right)\;.
\ee
Now since the left hand side is bounded from above by $g_0^{\frac{K}{a}}$, and from below by zero, the condition that there is no pole is
\be\label{eq:conditions}
\frac{K}{b}\frac{1}{\frac{\lambda_0}{g_0}+A_-}>0\quad \text{  or  }\quad \frac{K}{b}\frac{1}{\frac{\lambda_0}{g_0}+A_-}<-1\;.
\ee
For the first condition to be satisfied we must have 
\be\label{eq:ratio_condition}
\frac{\lambda_0}{g_0}>-A_-\;.
\ee
The second condition in \eqref{eq:conditions} can be satisfied only if the first is not, i.e., for  $\frac{\lambda_0}{g_0}+A_- < 0$. 
Upon multiplication of the second equation in (\ref{eq:conditions}) by $\frac{\lambda_0}{g_0}+A_-$ we would find the condition
\be
-A_->\frac{\lambda_0}{g_0}>-A_--\frac{K}{b} \; .
\ee
However, since $K/b>0$, the above equation is inconsistent because $-A_-<-a_--\frac{K}{b}$. Hence only the first condition 
makes sense and we find that the critical ratio of the bare couplings $\frac{\lambda_0}{g_0}$ is given by
\be\label{eq:critical_ratio}
r_c=-A_-=\frac{N_f+\sqrt{\left(N_f-180\right) N_f-540}+18}{6 \left(N_f+4\right)}\;,
\ee
and for $\lambda_0/g_0>r_c$ the transition is 2nd order. This is depicted in Fig.~\ref{fig:critical_ratio}.

\subsubsection{Solving the equations \eqref{eqs:RG_flow2}} 

We wish to solve the following RG equations
\begin{align}
&\frac{d g_1}{d\ell}= -a g_1^2\;,\\
&\frac{dg_2}{d\ell}= -b g_1^2-c g_1g_2\;.
\end{align}
Obviously $g_1(\ell)$ can be determined easily as
\be\label{app:g1}
g_1(\ell)= \frac{g_1(0)}{1+ag_1(0)\ell}\;.
\ee
Since $g_1(\ell)$ is monotonic, we can view $g_2$ as depending on $\ell$ through $g_1$, i.e.,
\be
\frac{dg_2}{d\ell}=\frac{dg_1}{d\ell}\frac{dg_2}{dg_1}=-a g_1^2 \frac{dg_2}{dg_1}\;,
\ee
where we used the RG equation for $g_1$. Now we see that the RG equation for $g_2$ becomes
\be
-ag_1^2 \frac{dg_2}{dg_1}=-b g_2^2-cg_1g_2  \; ,
\ee
or after rewriting
\be
\left(-ag_1^2\frac{d}{dg_1}+c g_1\right)g_2=-bg_2^2\;.
\ee
Now, replacing
\be
g_2(g_1)=g_1^{\frac{c}{a}}G(g_1)\;,
\ee
we find that $G(g_1)$ satisfies the equation
\be
ag_1^{2}G'(g_1)=bg_1^{\frac{c}{a}}G(g_1)^2\;,
\ee
which is solved by
\be
G(g_1)=-\frac{c-a}{b g_1^{{\frac{c}{a}-1}}+C}\;,
\ee
where $C$ is a constant. Hence $g_2(\ell)$ is given by
\be
g_2(\ell)=-(c-a)\frac{g_1(\ell)^{\frac{c}{a}}}{b g_1(\ell)^{{\frac{c}{a}-1}}+C} \; ,
\ee
where $g_1(\ell)$ is given by \eqref{app:g1}. By setting $\ell=0$ we find $C$ in terms of $g_1(0),g_2(0)$,
\be
C=- (c-a)\frac{g_1(0)^{\frac{c}{a}}}{g_2(0)}-b g_1(0)^{\frac{c}{a}-1}\;.
\ee

\end{appendix}

\vskip10mm
\bibliographystyle{utphys}
\bibliography{bibliography}

\providecommand{\href}[2]{#2}\begingroup\raggedright\begin{thebibliography}{10}

\bibitem{Haldane:1982rj}
F.~D.~M. Haldane, ``{Continuum dynamics of the 1-D Heisenberg antiferromagnetic
  identification with the O(3) nonlinear sigma model},'' {\em Phys. Lett. A}
  {\bf 93} (1983) 464--468.

\bibitem{Polyakov:1976fu}
A.~M. Polyakov, ``{Quark Confinement and Topology of Gauge Groups},'' {\em
  Nucl. Phys. B} {\bf 120} (1977) 429--458.

\bibitem{Senthil:2003eed}
T.~Senthil, A.~Vishwanath, L.~Balents, S.~Sachdev, and M.~P.~A. Fisher,
  ``{Deconfined Quantum Critical Points},'' {\em Science} {\bf 303} (2004),
  no.~5663 1490--1494, \href{http://xxx.lanl.gov/abs/cond-mat/0311326}{{\tt
  cond-mat/0311326}}.

\bibitem{Vishwanath:2003yjl}
A.~Vishwanath, L.~Balents, and T.~Senthil, ``{Quantum Criticality and
  Deconfinement in Phase Transitions Between Valence Bond Solids},'' {\em Phys.
  Rev. B} {\bf 69} (2004), no.~22 224416,
  \href{http://xxx.lanl.gov/abs/cond-mat/0311085}{{\tt cond-mat/0311085}}.

\bibitem{Pufu:2013vpa}
S.~S. Pufu, ``{Anomalous dimensions of monopole operators in three-dimensional
  quantum electrodynamics},'' {\em Phys. Rev. D} {\bf 89} (2014), no.~6 065016,
  \href{http://xxx.lanl.gov/abs/1303.6125}{{\tt 1303.6125}}.

\bibitem{Dyer:2015zha}
E.~Dyer, M.~Mezei, S.~S. Pufu, and S.~Sachdev, ``{Scaling dimensions of
  monopole operators in the $ \mathbb{C}{\mathrm{\mathbb{P}}}^{N_b-1} $ theory
  in 2 $+$ 1 dimensions},'' {\em JHEP} {\bf 06} (2015) 037,
  \href{http://xxx.lanl.gov/abs/1504.00368}{{\tt 1504.00368}}. [Erratum: JHEP
  03, 111 (2016)].

\bibitem{shao2016quantum}
H.~Shao, W.~Guo, and A.~W. Sandvik, ``Quantum criticality with two length
  scales,'' {\em Science} {\bf 352} (2016), no.~6282 213--216.

\bibitem{Montonen:1977sn}
C.~Montonen and D.~I. Olive, ``{Magnetic Monopoles as Gauge Particles?},'' {\em
  Phys. Lett. B} {\bf 72} (1977) 117--120.

\bibitem{Sulejmanpasic:2019ytl}
T.~Sulejmanpasic and C.~Gattringer, ``{Abelian gauge theories on the lattice:
  $\theta$-terms and compact gauge theory with(out) monopoles},'' {\em Nucl.
  Phys.} {\bf B943} (2019) 114616,
  \href{http://xxx.lanl.gov/abs/1901.02637}{{\tt 1901.02637}}.

\bibitem{Gattringer:2018dlw}
C.~Gattringer, D.~G{\"o}schl, and T.~Sulejmanpasic, ``{Dual simulation of the
  2d U(1) gauge Higgs model at topological angle $\theta = \pi\,$: Critical
  endpoint behavior},'' \href{http://xxx.lanl.gov/abs/1807.07793}{{\tt
  1807.07793}}.

\bibitem{Goschl:2018uma}
D.~G{\"o}schl, C.~Gattringer, and T.~Sulejmanpasic, ``{The critical endpoint in
  the 2d U(1) gauge-Higgs model at topological angle $\theta=\pi$},'' in {\em
  {36th International Symposium on Lattice Field Theory (Lattice 2018) East
  Lansing, MI, United States, July 22-28, 2018}}, 2018.
\newblock \href{http://xxx.lanl.gov/abs/1810.09671}{{\tt 1810.09671}}.

\bibitem{Sulejmanpasic:2020lyq}
T.~Sulejmanpasic, D.~G\"oschl, and C.~Gattringer, ``{First-Principles
  Simulations of 1+1D Quantum Field Theories at $\theta=\pi$ and Spin
  Chains},'' {\em Phys. Rev. Lett.} {\bf 125} (2020), no.~20 201602,
  \href{http://xxx.lanl.gov/abs/2007.06323}{{\tt 2007.06323}}.

\bibitem{Sulejmanpasic:2020ubo}
T.~Sulejmanpasic, ``{Ising model as a $U(1)$ lattice gauge theory with a
  $\theta$-term},'' {\em Phys. Rev. D} {\bf 103} (2021), no.~3 034512,
  \href{http://xxx.lanl.gov/abs/2009.13383}{{\tt 2009.13383}}.

\bibitem{Anosova:2021akr}
M.~Anosova, C.~Gattringer, N.~Iqbal, and T.~Sulejmanpasic, ``{Numerical
  simulation of self-dual U(1) lattice field theory with electric and magnetic
  matter},'' in {\em {38th International Symposium on Lattice Field Theory}},
  11, 2021.
\newblock \href{http://xxx.lanl.gov/abs/2111.02033}{{\tt 2111.02033}}.

\bibitem{Anosova:2022cjm}
M.~Anosova, C.~Gattringer, and T.~Sulejmanpasic, ``{Self-dual U(1) lattice
  field theory with a $\theta$-term},'' {\em Accepted for publication in JHEP}
  (1, 2022) \href{http://xxx.lanl.gov/abs/2201.09468}{{\tt 2201.09468}}.

\bibitem{Gorantla:2021svj}
P.~Gorantla, H.~T. Lam, N.~Seiberg, and S.-H. Shao, ``{A modified Villain
  formulation of fractons and other exotic theories},'' {\em J. Math. Phys.}
  {\bf 62} (2021), no.~10 102301,
  \href{http://xxx.lanl.gov/abs/2103.01257}{{\tt 2103.01257}}.

\bibitem{Choi:2021kmx}
Y.~Choi, C.~Cordova, P.-S. Hsin, H.~T. Lam, and S.-H. Shao, ``{Non-Invertible
  Duality Defects in 3+1 Dimensions},''
  \href{http://xxx.lanl.gov/abs/2111.01139}{{\tt 2111.01139}}.

\bibitem{Villain:1974ir}
J.~Villain, ``{Theory of one-dimensional and two-dimensional magnets with an
  easy magnetization plane. 2. The planar, classical, two-dimensional
  magnet},'' {\em J. Phys.(France)} {\bf 36} (1975) 581--590.

\bibitem{Elitzur:1979uv}
S.~Elitzur, R.~B. Pearson, and J.~Shigemitsu, ``{The Phase Structure of
  Discrete Abelian Spin and Gauge Systems},'' {\em Phys. Rev. D} {\bf 19}
  (1979) 3698.

\bibitem{Cardy:1981fd}
J.~L. Cardy, ``{Duality and the Theta Parameter in Abelian Lattice Models},''
  {\em Nucl. Phys. B} {\bf 205} (1982) 17--26.

\bibitem{Cardy:1981qy}
J.~L. Cardy and E.~Rabinovici, ``{Phase Structure of Z(p) Models in the
  Presence of a Theta Parameter},'' {\em Nucl. Phys. B} {\bf 205} (1982) 1--16.

\bibitem{Kapustin:2014gua}
A.~Kapustin and N.~Seiberg, ``{Coupling a QFT to a TQFT and Duality},'' {\em
  JHEP} {\bf 04} (2014) 001, \href{http://xxx.lanl.gov/abs/1401.0740}{{\tt
  1401.0740}}.

\bibitem{Kaidi:2021xfk}
J.~Kaidi, K.~Ohmori, and Y.~Zheng, ``{Kramers-Wannier-like duality defects in
  (3 + 1)d gauge theories},'' \href{http://xxx.lanl.gov/abs/2111.01141}{{\tt
  2111.01141}}.

\bibitem{Kravec:2013pua}
S.~M. Kravec and J.~McGreevy, ``{A gauge theory generalization of the
  fermion-doubling theorem},'' {\em Phys. Rev. Lett.} {\bf 111} (2013) 161603,
  \href{http://xxx.lanl.gov/abs/1306.3992}{{\tt 1306.3992}}.

\bibitem{Gaiotto:2014kfa}
D.~Gaiotto, A.~Kapustin, N.~Seiberg, and B.~Willett, ``{Generalized Global
  Symmetries},'' {\em JHEP} {\bf 02} (2015) 172,
  \href{http://xxx.lanl.gov/abs/1412.5148}{{\tt 1412.5148}}.

\bibitem{Fradkin:1978dv}
E.~H. Fradkin and S.~H. Shenker, ``{Phase Diagrams of Lattice Gauge Theories
  with Higgs Fields},'' {\em Phys. Rev. D} {\bf 19} (1979) 3682--3697.

\bibitem{Coleman:1973jx}
S.~R. Coleman and E.~J. Weinberg, ``{Radiative Corrections as the Origin of
  Spontaneous Symmetry Breaking},'' {\em Phys. Rev. D} {\bf 7} (1973)
  1888--1910.

\bibitem{kolnberger1990critical}
S.~Kolnberger and R.~Folk, ``Critical fluctuations in superconductors,'' {\em
  Physical Review B} {\bf 41} (1990), no.~7 4083.

\bibitem{Folk:1998yt}
R.~Folk and Y.~Holovatch, ``{Critical fluctuations in normal to superconducting
  transition},'' in {\em {1st Winter Workshop on Cooperative Phenomena in
  Condensed Matter}}, 7, 1998.
\newblock \href{http://xxx.lanl.gov/abs/cond-mat/9807421}{{\tt
  cond-mat/9807421}}.

\bibitem{Ihrig:2019kfv}
B.~Ihrig, N.~Zerf, P.~Marquard, I.~F. Herbut, and M.~M. Scherer, ``{Abelian
  Higgs model at four loops, fixed-point collision and deconfined
  criticality},'' {\em Phys. Rev. B} {\bf 100} (2019), no.~13 134507,
  \href{http://xxx.lanl.gov/abs/1907.08140}{{\tt 1907.08140}}.

\bibitem{Hofman:2018lfz}
D.~M. Hofman and N.~Iqbal, ``{Goldstone modes and photonization for higher form
  symmetries},'' {\em SciPost Phys.} {\bf 6} (2019), no.~1 006,
  \href{http://xxx.lanl.gov/abs/1802.09512}{{\tt 1802.09512}}.

\bibitem{Kerler:1996cr}
W.~Kerler, C.~Rebbi, and A.~Weber, ``{Critical properties and monopoles in U(1)
  lattice gauge theory},'' {\em Phys. Lett. B} {\bf 392} (1997) 438--443,
  \href{http://xxx.lanl.gov/abs/hep-lat/9612001}{{\tt hep-lat/9612001}}.

\bibitem{Damm:1997vd}
G.~Damm and W.~Kerler, ``{Critical exponents in U(1) lattice gauge theory with
  a monopole term},'' {\em Nucl. Phys. B Proc. Suppl.} {\bf 63} (1998)
  703--705, \href{http://xxx.lanl.gov/abs/hep-lat/9709061}{{\tt
  hep-lat/9709061}}.

\bibitem{Somoza:2020jkq}
A.~M. Somoza, P.~Serna, and A.~Nahum, ``{Self-Dual Criticality in
  Three-Dimensional Z2 Gauge Theory with Matter},'' {\em Phys. Rev. X} {\bf 11}
  (2021), no.~4 041008, \href{http://xxx.lanl.gov/abs/2012.15845}{{\tt
  2012.15845}}.

\bibitem{Mercado:2013yta}
Y.~Delgado~Mercado, C.~Gattringer, and A.~Schmidt, ``{Surface worm algorithm
  for abelian Gauge-Higgs systems on the lattice},'' {\em Comput. Phys.
  Commun.} {\bf 184} (2013) 1535--1546,
  \href{http://xxx.lanl.gov/abs/1211.3436}{{\tt 1211.3436}}.

\bibitem{Mercado:2013ola}
Y.~Delgado~Mercado, C.~Gattringer, and A.~Schmidt, ``{Dual Lattice Simulation
  of the Abelian Gauge-Higgs Model at Finite Density: An Exploratory Proof of
  Concept Study},'' {\em Phys. Rev. Lett.} {\bf 111} (2013) 141601,
  \href{http://xxx.lanl.gov/abs/1307.6120}{{\tt 1307.6120}}.

\bibitem{Halperin:1973jh}
B.~I. Halperin, T.~C. Lubensky, and S.-K. Ma, ``{First order phase transitions
  in superconductors and smectic A liquid crystals},'' {\em Phys. Rev. Lett.}
  {\bf 32} (1974) 292--295.

\end{thebibliography}\endgroup

\end{document}